\documentclass{ptephy_v1}

\preprintnumber{A} 

\usepackage{amsmath}
\usepackage{mathrsfs} 
\usepackage{hyperref}

\usepackage{graphics}
\usepackage{url}

\DeclareMathOperator{\sinc}{sinc}

\newcommand{\ii}{\textrm{i}}

\begin{document}

\title{A back-linked Fabry-Perot interferometer for space-borne gravitational wave observations}


\author[1]{Kiwamu Izumi}
\affil{Institute of Space and Astronautical Science, Japan Aerospace Exploration Agency, \\3-1-1 Yoshinodai, Chuo-ku, Sagamihara, Kanagawa, Japan 252-5210\email{kiwamu@astro.isas.jaxa.jp}}

\author[2]{Masa-Katsu Fujimoto}
\affil{National Astronomical Observatory of Japan, 2-21-1 Osawa, Mitaka, Tokyo, Japan 181-0015}

\begin{abstract}%
Direct observations of gravitational waves at frequencies below 10~Hz will play crucial roles for fully exploiting the potential of gravitational wave astronomy. One approach to pursue this direction is the utilization of laser interferometers equipped with the Fabry-Perot optical cavities in space. However, a number of challenges lie in this path practically. In particular, the implementation of precision control for the cavity lengths and the suppression of laser phase noises may prevent a practical detector design. To circumvent such difficulties, we propose a new interferometer topology, named the back-linked Fabry-Perot interferometer, where the precision length controls are not required and an offline subtraction scheme for laser phase noises is readily applicable. This article presents the principle idea and the associated sensitivity analyses. Despite additional noises, a strain sensitivity of $7\times10^{-23}\,\textrm{Hz}^{-1/2}$ may be attainable in the deci-Hertz band. Several technological developments and studies must be carried out to pave the way forward for the implementation. 
\end{abstract}

\subjectindex{xxxx, xxx}

\maketitle

\section{Introduction} \label{sec:intro}
Following the first direct observation of gravitational waves~\cite{Abbott:2016ki}, the ground-based laser-interferometric gravitational wave detectors~\cite{aligo, avirgo} continue to provide unique observations on the dark and relativistic side of our Universe. A number of remarkable astrophysical contributions have been brought by these detectors, including the observational evidences of the r-process~\cite{Abbott_2017}, determinations of the Hubble constant combined with the electromagnetic observations~\cite{Abbott:2017tj,Hotokezaka:2019ui} and estimations of the black hole population~\cite{Abbott_2019}. The current and future ground-based detectors~\cite{Abbott:2018vl,GEOlatest} are sensitive at frequencies above typically 10~Hz, corresponding to probing the low-mass regime of the binary systems with the total mass up to $\sim100$~$M_\odot$.

To fully exploit the potential of gravitational wave astronomy, the frequency band below 10~Hz must be probed. Such observations will be crucial not only for studying the high-mass regime of the binary systems including super-massive and intermediate-mass black holes, but also for exploring other types of astrophysically interesting sources such as relic gravitational wave backgrounds~\cite{1975JETP...40..409G}. In addition, the observations of stellar mass binary systems at low frequencies can issue the forewarnings to the ground-based detectors as well as the electro-magnetic telescopes. 

In practice, probing the band below 10~Hz requires observations in space in order to escape from terrestrial seismic noise which severely limits the sensitivity of the ground-based ones at low frequencies~\cite{RevModPhys.86.121,Martynov:2016gf}. For the reason, several space laser interferometers have been proposed to date, covering the frequency range approximately from $10^{-4}$ to 10~Hz, including LISA~\cite{amaroseoane2017laser}, TianQin~\cite{Luo_2016}, Taiji~\cite{10.1093/nsr/nwx116}, DECIGO~\cite{kawamura2006japanese}, BBO~\cite{Harry:2006fz} and TianGO~\cite{kuns2019astrophysics}. An intriguing design choice for such space interferometry is the use of the Fabry-Perot optical cavity, as is the case for DECIGO, to enhance the sensitivity.

However, the Fabry-Perot cavities with a cavity length longer than 100~km in orbit seems rather challenging. In a naive design, the cavity lengths would have to be controlled at a precision level better than $10^{-9}$~m to maintain the resonances. This implies an implementation of formation flying with an unprecedentedly high precision for a distance scale of over 100~km. A straightforward workaround is to control instead the frequency of the lasers while leaving the cavity lengths uncontrolled. However, this approach makes the sensitivity vulnerable to laser phase noise because laser noises will not be common-mode rejected any more. 
For instance, reaching gravitational wave sensitivity of $10^{-23}$~Hz$^{1/2}$ at around 0.1~Hz requires phase stability of $\sim10^{-8}$~rad/Hz$^{1/2}$ which is much better than the typical free-running noise of NPRO lasers by a factor of more than $10^{12}$. There has not been a viable solution to this conundrum so far.

In this article, we present a new laser interferometer topology, named the {\it back-linked Fabry-Perot interferometer} or BLFP interferometer in short, which solves the conundrum. The BLFP interferometer keeps each Fabry-Perot cavity at a resonance by controlling the lasers' frequencies and yet is immune against laser phase noises because it offers the implementation of a new offline noise subtraction scheme similarly to time-delay interferometry~\cite{PhysRevD.59.102003}. Therefore, the BLFP interferometer relaxes the demand for high-precision formation flying and the requirement on laser phase noises simultaneously. We present the principle idea and discuss an achievable sensitivity through a set of analyses.

\begin{figure}[t]
	\begin{center}
			\includegraphics[width=0.5\columnwidth]{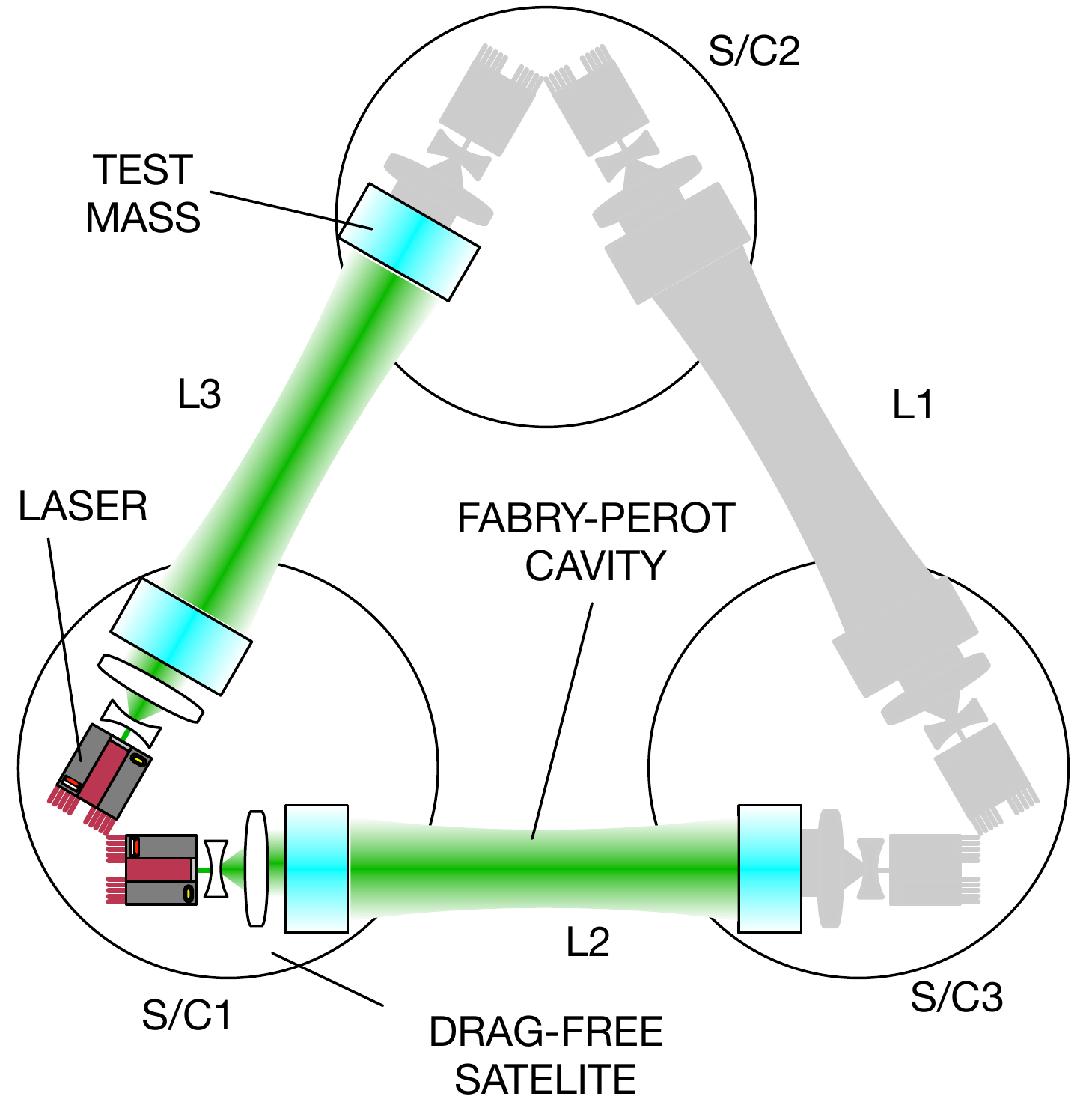}
	\end{center}
\caption{A schematic view of the constellation for the BLFP interferometer. One of three interferometers, formed by Spacecraft 1 with Fabry-Perot cavities $L_2$ and $L_3$, is highlighted in color while the other two are shaded in gray for illustrative purpose. S/C stands for spacecraft.\label{fig:constellation}}
\end{figure}

The organization of the article is as follows. Section~\ref{sec:backlink} describes the interferometer setup and the working principle qualitatively. Section~\ref{sec:offline} proposes the scheme for subtracting laser phase noise. Section~\ref{sec:trans} discusses various transfer functions for both gravitational waves and selected noise sources. Section~\ref{sec:sens} discusses a possible strain sensitivity and its astrophysical implications. In section~\ref{sec:discussion}, we discuss technological developments and studies to be addressed. Finally, we conclude our study in section~\ref{sec:conclusion}.

\section{The back-linked Fabry-Perot interferometer} \label{sec:backlink}
\subsection{The configuration}
\begin{figure}[t]
	\begin{center}
			\includegraphics[width=0.6\columnwidth]{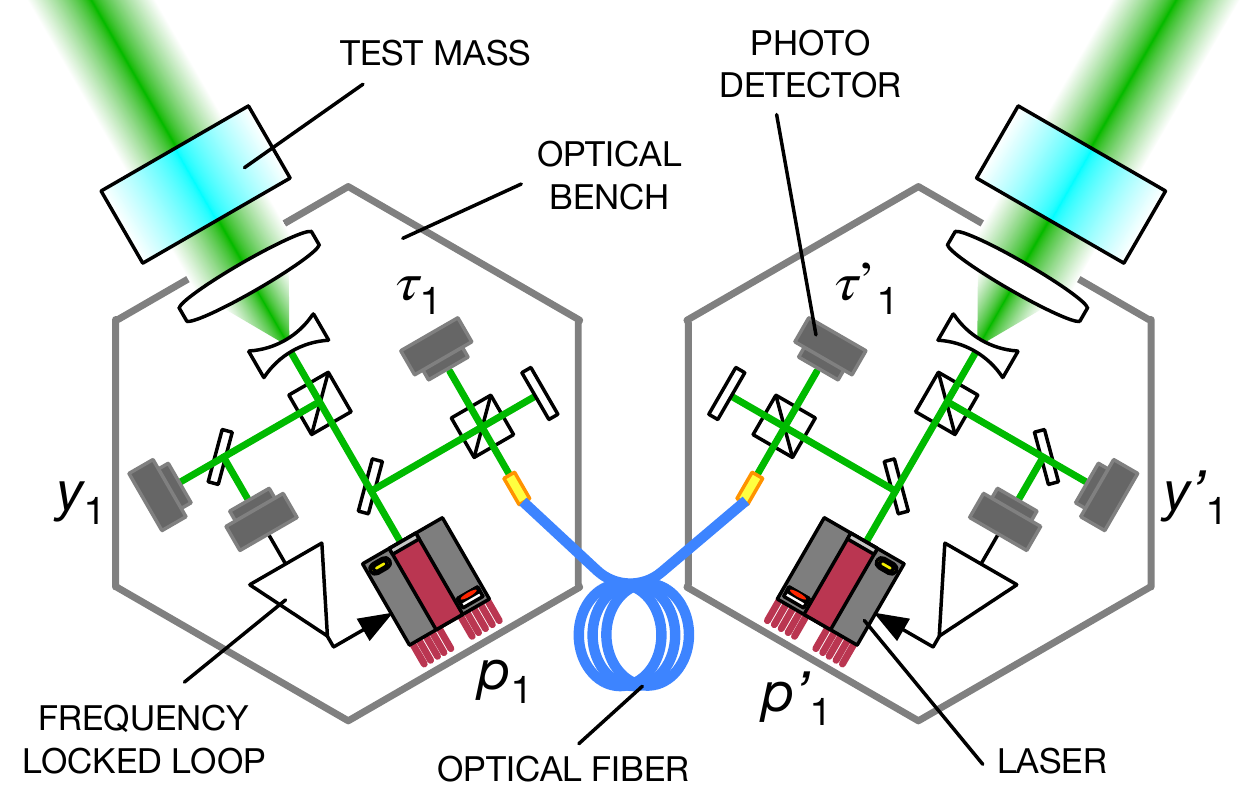}
	\end{center}
\caption{A schematic view of the optical setup in Spacecraft 1.\label{fig:BLFP}}
\end{figure}

The constellation and the optical setup of the BLFP interferometer are schematically sketched in Figures~\ref{fig:constellation} and \ref{fig:BLFP}, respectively. The constellation consists of three identical spacecraft forming an almost equilateral triangle. Each spacecraft houses two floating test masses serving as the references for measuring gravitational waves. The test masses also serve as partially reflective mirrors, forming three Fabry-Perot cavities in total. Each of the Fabry-Perot cavities is illuminated by lasers from both sides. The cavities enhance the effect of incoming gravitational waves by elongating the interaction time of the laser light with the gravitational waves, leading to a better sensitivity compared to the Michelson interferometer~\cite{drever:1991}. The position of a spacecraft relative to the test masses is drag-free controlled in order to shield the test masses from external disturbances as was demonstrated in LISA Pathfinder~\cite{Armano:2016hp}. 

The overall configuration is similar to that proposed for DECIGO except that each spacecraft explicitly carries two independent lasers in the BLFP. Furthermore, the two lasers onboard a spacecraft are optically communicated through an optical fiber which we call the {\it back-link} similarly to LISA~\cite{Steier:2009cn,Fleddermann:2018ej,Isleif:2018fq}. We note that the back-link does not have to be a fiber-coupled optical path and can be a free-space one.

Three spacecraft are enumerated in clockwise as shown in Figure~\ref{fig:constellation}, running from 1 through 3. The cavity lengths $L_j$ are named in such a way that the number in the subscript corresponds to that of the opposite spacecraft. In the $j$-th spacecraft, the relevant quantities are subscripted as $A_j, A'_j$ where the non-primed and primed quantities represent those associated with the left and right optical benches, respectively, as illustrated in Figure~\ref{fig:BLFP}. This cyclic convention is intentionally chosen to make our discussion coherent with those for time-delay interferometry~\cite{Tinto:2014ki} such that confusions are minimized.

\subsection{The working principle \label{sec:wp}}
The frequency of Laser 1 on the left optical bench in Spacecraft 1 is locked to the length of the corresponding Fabry-Perot cavity $L_3$ such that the cavity is resonant for the laser light. Locking is achieved by using a control system feeding the cavity-length signals back to the laser frequency. Unlike LISA or similar concepts, such a frequency-locked loop is necessary in order to operate the Fabry-Perot interferometer as a linear measurement apparatus. The same is true for the other laser, namely Laser 1', the frequency of which is locked to $L_2$. 

The frequency response of a Fabry-Perot cavity to the laser frequency exhibits unstable null points at high frequencies. The lowest null point occurs at 1.5~kHz for a cavity length of 100~km, corresponding to the free spectral range of the cavity. We assume, throughout the paper, that the control bandwidths of the frequency-locked loops are smaller than the free spectral range so that the implementation of the control system does not require an elaborate scheme such as those studied for LISA arm locking~\cite{SHEARD20039,PhysRevD.78.082001}.

With the aid of the frequency-locked loop, the frequency of Laser 1 satisfies the resonance condition,
	\begin{equation}
		2 \pi n = \frac{4 \pi \nu_{1} L_3}{c},
	\end{equation}
where $n$ is an arbitrary integer, and where $\nu_1$ and $c$ are the laser frequency and the speed of light, respectively. This relation silently assumes that the frequency-locked loop has an infinitely large gain. This assumption is not realistic. In fact, the removal of the assumption will reveal the serious issue of laser phase noise contaminating the observatory sensitivity. We will return to this point in Section~\ref{sec:offline}.

Taking the derivative of both sides in the equation above, one can find
	\begin{equation}
		\frac{\Delta \nu_1}{\nu_1} = -\frac{\Delta L_3}{L_3}. \label{eq:dnudL}
	\end{equation}
Thus, length variation $\Delta L_3$ is linearly imprinted onto a shift in the laser frequency $\Delta \nu_1$ via the frequency-locked loop. The right hand side represents a strain, roughly equivalent to the amplitude of gravitational waves in the long wavelength limit. This argument holds for the other Fabry-Perot cavity, $L_2$, which is resonant for Laser 1'.

Subsequently, the back-link path offers an optical heterodyne measurement where the fields of the two lasers interfere, leading to a sinusoidal modulation in the laser intensity received by the photodetectors $\tau_1$ and $\tau'_1$. Since the modulation frequency is equal to the difference between the frequencies of two lasers, its variation can be expressed as $\Delta f_\textrm{hetero} = \Delta \nu_1 -\Delta \nu'_1$. Using equation~(\ref{eq:dnudL}) and the relation $\Delta L_j/L_j\sim h_j$ with $h_j$ being the amplitude of gravitational waves passing through $L_j$, one can express variations of the heterodyne frequency as
	\begin{equation}
		\Delta f_\textrm{hetero} =  \nu_1 \left( h_3 - h_2 \right),
	\end{equation}
where we have assumed $\nu_1 \approx \nu'_1$. Therefore, the back-link heterodyne provides measurements of the incoming gravitational waves. This is the working principle of the BLFP interferometer. A more rigorous analysis is presented in Section~\ref{sec:trans}. A similar concept has been successfully implemented for the initial length control of the Fabry-Perot cavities in the ground-based interferometers~\cite{Izumi:2012wv,Mullavey:2012wv,Staley:2014hg,Akutsu_2020}.

\subsection{Advantages\label{sec:adv}}
A first advantage is the tractability in achieving the resonances of the Fabry-Perot cavities. Since each Fabry-Perot cavity is optically coupled to two laser fields where each field is incident on each end of the cavity, there are six resonance conditions to meet for a single constellation. The BLFP interferometer can simultaneously achieve all six resonances regardless of how long the absolute lengths of the cavities are because of the use of six independent lasers.

However, in contrast, this is not the case for the conventional configurations which use three lasers with the field from each laser split into two fields by an additional beam splitter onboard each spacecraft. In fact, it requires the absolute lengths of the cavities to be a combination of certain values. This restriction is a consequence of the fact that the frequency of a laser is interrelated with the other two via the active controls acting on both cavity lengths and laser frequencies~\cite{knagano2020}. So for the reason, the conventional configuration will likely need an iterative process to arrange the absolute cavity lengths in orbit, whereas the BLFP does not. Thus, the mission risk associated with the resonance acquisition can be reduced with the BLFP interferometer.

Another advantage is that the BLFP interferometer permits the cavity lengths to evolve as functions of time as long as the frequency-locked loops are maintained. This is not the case for the conventional one which actively controls the cavity lengths to be constant at a precision level better than $10^{-9}$~m for maintaining the resonances of the Fabry-Perot cavities. This is equivalent to achieving high-precision formation flying in orbit. In this case, the design and implementation of the orbit for the constellation can be even more involved. Therefore, the BLFP reduces the complexity in the orbit design and implementation, resulting in an increased mission feasibility.

On the other hand, the biggest conundrum in adopting the BLFP configuration is that it would make the noise contribution of laser phase noises much more severe. This is because the BLFP interferometer can not provide common-mode rejection for laser noises as the noises of a laser is incoherent to that of the others. We show that the conundrum can be resolved because the BLFP interferometer allows for a relatively easy implementation of offline subtraction of laser phase noise. Therefore, it is much less vulnerable against laser phase noise which could be predominant if unaddressed. We discuss the subtraction scheme in detail in the following section.

\section{Subtracting laser phase noise} \label{sec:offline}
Observations of gravitational waves as small as $10^{-23}\,\text{Hz}^{-1/2}$ require extremely small laser frequency noise (or equivalently phase noise) at a level of $10^{-9}\,\text{Hz}/\sqrt{\text{Hz}}$ according to equation~(\ref{eq:dnudL}) for lasers with $\nu=5.8\times10^{14}$~Hz. Even if we assume an excellent stability of approximately 0.1~Hz/$\sqrt{\text{Hz}}$ that has been available on a laboratory scale~\cite{Numata:2004de}, an improvement by eight orders of magnitude is necessary.

Even though the frequency-locked loops can suppress the phase noise further by perhaps a factor of $10^4$ at around 1~Hz, the residual noise will be still too large. For the reason, a novel idea of reducing laser phase noise even further by a factor of $10^4$ has been sought. Inspired by the excellent noise subtraction scheme in time-delay interferometry~\cite{Armstrong_1999}, we developed an offline noise subtraction scheme which solves this issue. This section is devoted for describing the subtraction scheme in detail.

\medskip
\subsection{Fabry-Perot cavity as a phase detector}
\begin{figure}[t]
	\begin{center}
			\includegraphics[width=0.6\columnwidth]{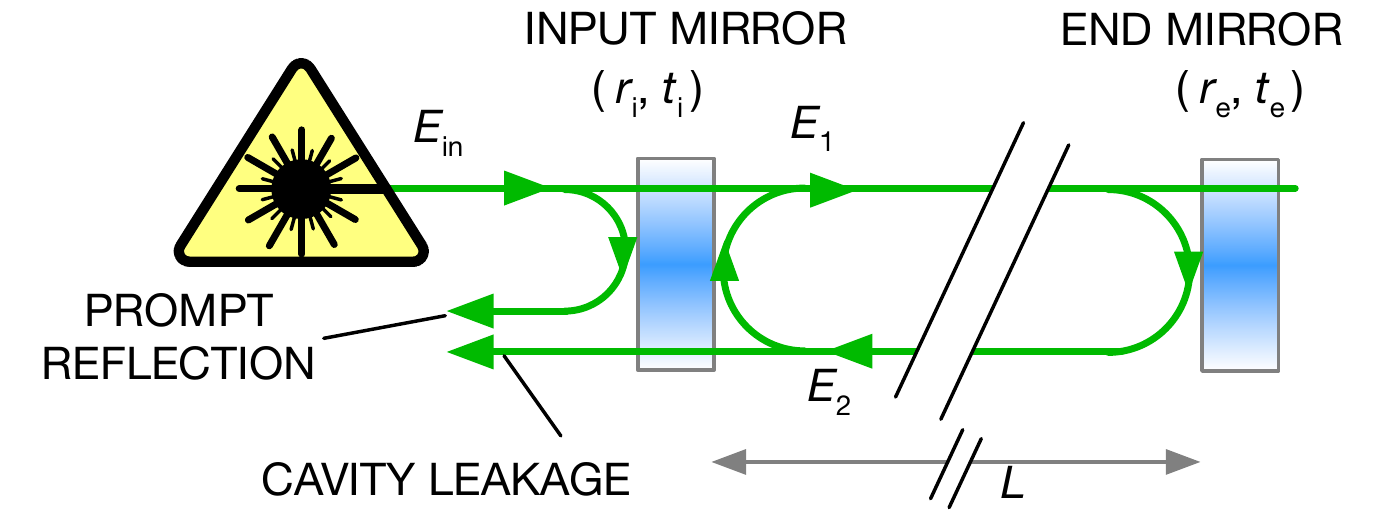}
	\end{center}
\caption{A schematic view of a Fabry-Perot cavity.\label{fig:virtualFP}}
\end{figure}

We show that the Fabry-Perot cavity can be treated as a linear sensor for the phase of the input laser light $p(t)$. As will be discussed shortly, its response can be expressed by
	\begin{equation}
		y(t) =  p(t) - \ell \ast  p, \label{eq:ytfp}
	\end{equation}
where the asterisk denotes the convolution integral and $\ell $ is a linear operator with its Fourier transform $\mathscr L$ given by
	\begin{equation}
		 \mathscr L\left( f\right)= \frac{\left(1-r_\text{i}r_\text{e} \right)e^{-4\textrm{i}\pi f T }}{1-r_\text{i}r_\text{e} e^{-4\textrm{i}\pi f T }} , \label{eq:L}
	\end{equation}
where $T\equiv L/c$ is a single-trip time for the cavity distance $L$, and where $r_\textrm{i}$ and $r_\textrm{e}$ are the amplitude reflectivity of the input and end mirrors, respectively (see Figure~\ref{fig:virtualFP}). This response function is applicable for the standard readout scheme, namely the Pound-Drever-Hall scheme~\cite{Drever:1983wu}.

It is quite interesting that the Fabry-Perot response (\ref{eq:ytfp}) resembles that of the Michelson interferometer with the distance of one of the arms set to zero. Such a response is the most important building block in time-delay interferometry and shows the following response function~\cite{Tinto:2014ki},
	\begin{equation}
		y_\textrm{mi} (t) = p(t) - \mathcal D(2T)\, p(t), \label{eq:ymi}
	\end{equation}
where $\mathcal D(\tau)$ is the time delay operator that exerts a time-delay shift so that $\mathcal D (\tau) X(t) \equiv X(t-\tau)$. In the case of the Fabry-Perot cavity, the operator $\mathcal D$ is replaced by $\ell$. As discussed later, equations~(\ref{eq:ytfp}) and (\ref{eq:L}) play the centric role in our noise subtraction scheme.

In what follows, we derive equations~(\ref{eq:ytfp}) and (\ref{eq:L}) based on a simplified argument where the effect of the modulation and demodulation processes are ignored. A rigorous derivation for the case of the Pound-Drever-Hall scheme is given in appendix~\ref{sec:pdh}.

Let us suppose that a Fabry-Perot cavity with the cavity length of $L$ is illuminated by a laser field as shown in Figure~\ref{fig:virtualFP}. The intracavity field leaving the input mirror $E_1$ and that arriving back at the input mirror after a round trip $E_2$ can be expressed, respectively, as
	\begin{eqnarray}
		E_1(t)  &=& t_\text{i} E_\text{in} (t) + r_\text{i} E_2 (t),\\
		E_2(t) &=& r_\textrm{e}  E_1(t-2T) e^{-2 \textrm{i}kL}, 
	\end{eqnarray}
where $E_\textrm{in}$ represents the input field and where $k$ is the wave number. The Fabry-Perot cavity is hereafter assumed to be kept at a resonance by some means so that $kL = n\pi$ with $n$ an integer. Eliminating $E_1$ by combining the two equations, one gets,
	\begin{equation}
		E_2(t) - r_\text{i} r_\text{e} E_2(t-2T) = t_\text{i} r_\text{e} E_\text{in}(t-2T).\label{eq:intraE}
	\end{equation}

It is convenient to switch the notation of the fields to the exponential format, so that
	\begin{equation}
		E_\textrm{in} = A_0e^{\textrm{i} p(t)}\quad\textrm{and}\quad E_2(t) = Ae^{\textrm{i}\Psi(t)}.
	\end{equation}
Plugging these into equation~(\ref{eq:intraE}), expanding the exponential terms to the first order and evaluating the imaginary part, one gets
	\begin{equation}
		\Psi(t) = r_\textrm{i} r_\textrm{e} \Psi (t-2T) + \left(1-r_\textrm{i} r_\textrm{e} \right) p(t-2T),
	\end{equation}
where we used $A=A_0 t_\textrm{i}r_\textrm{e}/(1-r_\text{i}r_\text{e})$ which had been obtained by evaluating the real part of the same equation.
We note that the extra phase components arose by slow length drift are absorbed into the locking condition, $kL=n\pi$. Therefore, the instantaneous phases $p$ and $\Psi$ can still be assumed to be small.

The above can be rewritten by using a summation as
	\begin{equation}
		\Psi(t) =  \left(1-r_\textrm{i} r_\textrm{e} \right) \sum_{m=1}^\infty \left(r_\textrm{i} r_\textrm{e}\right)^{m-1} p(t-2mT),
	\end{equation}	
where we have used the relation $r_\text{i}r_\text{e} < 1$ to remove the term proportional to $\Psi (t-2mT)$.
Introducing the Fourier transform of the phase,
	\begin{equation}
		p(t) = \int^\infty_{-\infty} \tilde p\left(\omega\right) e^{\textrm{i} \omega t} d\omega, \label{eq:fourier}
	\end{equation}
and executing the summation, one can arrive at
	\begin{equation}
		\Psi (t) = \int^\infty_{-\infty} \mathscr L (\omega) \tilde p\left( \omega \right) e^{\textrm{i}\omega t} d\omega.
	\end{equation}
This means that the phase of the leakage field $E_2$ conveys the phase information of the input light with the cavity effect imposed i.e., $\Psi (t) = \ell \ast p$.

In the case of the Pound-Drever-Hall readout scheme and similar ones, the phase of the leakage field $\Psi$ is then compared against the phase of the prompt reflection (see Figure~\ref{fig:virtualFP}). Since the prompt reflection does not acquire the cavity effect $\mathscr L$, the response should be in the form of
	\begin{equation}
		y(t) = p(t) - \int^\infty_{-\infty}  \mathscr L (\omega) \tilde p\left( \omega\right) e^{\textrm{i} \omega t}d\omega.
	\end{equation}
This is equivalent to equation~(\ref{eq:ytfp}). In the frequency domain, this can be expressed as
	\begin{equation}
		\tilde {y}(f) = \left [ 1 - \mathscr L \right] \tilde p(f). \label{eq:yfd}
	\end{equation}
Additionally, a few relevant characteristics of the response are reviewed in Appendix~\ref{sec:char}.
\begin{figure}[t]
	\begin{center}
			\includegraphics[width=0.6\columnwidth]{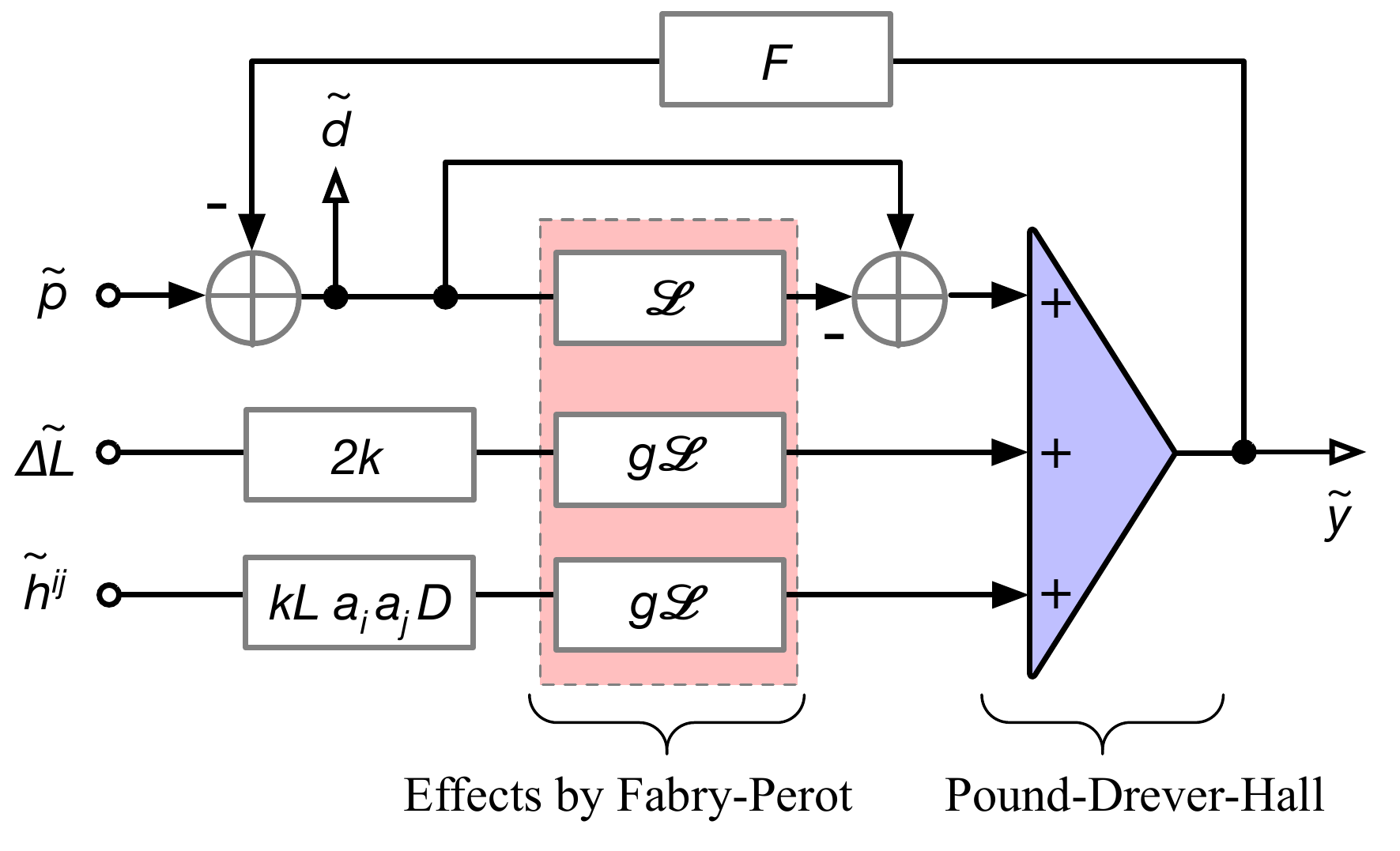}
	\end{center}
\caption{A block diagram representation of the frequency locked-loop for a laser in the frequency domain. The system is sensitive to laser phase noise $p$, cavity displacement $\Delta L$ and gravitational waves $h$. All of these inputs experience the effect of the Fabry-Perot cavity (highlighted by a red-shaded box) and additively emerge as Pound-Drever-Hall signals (the triangular-shaped summation symbol). Two signal extraction points are placed. One is right after the summation, $y$,  and the other after the feedback point, $d$. \label{fig:loop}}
\end{figure}

\subsection{The subtraction scheme}
A derivation and the description of our subtraction scheme for laser phase noise are introduced in this subsection. As opposed to the existing time-delay interferometry techniques, the scheme does not use a time-delay operator. It instead uses a set of the filtering operators, $\ell$.

Even though the BLFP interferometer employs a set of the frequency-locked loops as shown in figure~\ref{fig:loop}, the subtraction scheme can be derived without considering the effects from the frequency-locked loops. This is equivalent to removing the block denoted with $F$ in figure~\ref{fig:loop} with the assumption that the Fabry-Perot cavity stays at a resonance by some means. For the reason, the derivation here does not consider the loop effects. The subtraction scheme is further validated with the frequency-locked loops fully incorporated in Section~\ref{sec:33}.

For a laser on the $j$-th spacecraft, the phase measurement derived by the Pound-Drever-Hall readout (or similar methods) $y_j$ gives,
	\begin{equation}
		\tilde y_j (f) = \left ( 1 - \mathscr L_j \right) \tilde p_j + \mathscr H_k \tilde h_k(f), \label{eq:pdh}
	\end{equation}
where $h_k$ and $\mathscr H_k$ are the amplitude of incoming gravitational waves passing through $L_k$ and its transfer function, respectively. We will carefully analyze $\mathscr H_k$ in Section~\ref{sec:trans} and leave it as it is in this particular discussion. We do not deal with length fluctuations of the Fabry-Perot cavity $\Delta L$ for now for simplicity. At this point, it is clear that gravitational wave signals are contaminated by laser phase noise because one can not distinguish whether variations in $y$ is a consequence of gravitational waves $h$ or phase noise $p$.

Suppose one is able to perform another phase measurement $d_j$ monitoring the laser phase stability i.e., $d_j = p_j$, before the laser light enters the Fabry-Perot cavity. Now, linearly combining the recorded data of these two phase measurements, one can construct a new data set which completely removes laser phase noise without diminishing gravitational wave signals,
	\begin{equation}
		\tilde W_j(f) = \tilde y_j - (1-\mathscr L_j)\,\tilde d_j(f) = \mathscr H_k \tilde h_k(f). \label{eq:singlesubtraction}
	\end{equation}
This is the essence of the subtraction of laser phase noise in our scheme. This offline synthesis can be implemented either in the frequency domain or time domain, although the frequency domain implementation may not be accurate enough due to the effect from the finite-length Fourier transforms~\cite{PhysRevD.59.102003}.

In reality, one has to prepare a monitor measurement equivalent to the $d$ signals in order to implement our phase-noise subtraction scheme. Fortunately, the optical heterodyne measurements provided by the back-link $\tau_j$ and $\tau'_j$ are able to do so.
The two back-link heterodyne measurements in Spacecraft 1 can be expressed as
	\begin{eqnarray}
		\tau_1 (t)= d_1 - d'_1 + \mu_1, \label{eq:tau}\\
		\tau'_1 (t)= d'_1 -d_1 + \mu_1, \label{eq:taud}
	\end{eqnarray}
where $\mu_1$ represents common phase noise picked up by the light as they propagate through the optical fiber. Ideally, such noise should be common to both measurements and therefore removable to the level where other non-reciprocal phase noise stand out such as back-scattering noise~\cite{Isleif:2018fq}. We hereafter assume the non-reciprocal phase noises to be sufficiently small.

Taking the two Fabry-Perot cavities into account, we can now construct the following data set,
	\begin{equation}
	\begin{aligned}
		X_1(t) =& (1-\ell'_1\ast) y_1(t) - (1-\ell_1\ast) y'_1(t)\\
			& -(1-\ell_1\ast ) (1-\ell'_1\ast) \frac{\tau_1(t) - \tau'_1(t)}{2}, \label{eq:postproccess}
	\end{aligned}
	\end{equation}
where laser phase noises $p_1$ and $p'_1$ are completely cancelled out.
This is the complete recipe for our phase noise subtraction scheme.

To verify the subtraction, plugging equations~(\ref{eq:tau}) and (\ref{eq:taud}) into the above in the frequency domain, one gets
	\begin{equation}
	\begin{aligned}
		\tilde X_1(f) =& (1-\mathscr L'_1 )\left[  \tilde y_1-\left( 1-\mathscr L_1\right) \tilde d_1\right]\\
			&- (1-\mathscr L_1) \left[ \tilde y'_1 -(1-\mathscr L'_1 ) \tilde d'_1 \right]. \label{eq:X11}
	\end{aligned}
	\end{equation}
Using the subtraction relation for the individual lasers (\ref{eq:singlesubtraction}), one arrives at
	\begin{equation}
		\tilde X_1 (f) = (1-\mathscr L'_1 )\mathscr H_3 \tilde h_3 -(1-\mathscr L_1 ) \mathscr H'_2 \tilde h'_2. \label{eq:xresult}
	\end{equation}
Therefore, this scheme indeed removes laser phase noises of Lasers 1 and 1'.
We note that one needs to calibrate the synthesized output $X_1$ by essentially multiplying the inverse of the cavity response $(1-\mathscr L)$ to obtain the strain amplitude. Discussions on the signal-to-noise ratio are given in Sections~\ref{sec:trans} and \ref{sec:sens}.

\subsection{Incorporating the frequency-locked loops \label{sec:33}}
We now fully incorporate the frequency-locked loops to further validate the subtraction scheme. For a laser on the $j$-th spacecraft, the Pound-Drever-Hall readout is suppressed by the active feedback control, resulting in,
	\begin{equation}
		\tilde y_j (f)= \frac{1-\mathscr L_j}{1+G_j}\,\tilde p_j (f) + \frac{\mathscr H_k}{1+G_j}\,\tilde h_k(f), \label{eq:yf}
	\end{equation}
where $G_j$ is the open-loop gain defined by $G_j\equiv F_j(1-\mathscr  L_j)$ with $F_j$ being a linear servo filter as shown in Figure~\ref{fig:loop}. 
As opposed to the simple case~(\ref{eq:pdh}), both laser phase noise and gravitational wave signals are suppressed by the frequency-locked loop by a factor of $(1+G_j)^{-1}$ where typically $|G_j| \gg 1$ at frequencies below the control bandwidth. 

Similarly, the phase stability monitor $d$ is altered as,
	\begin{equation}
		\tilde d_j(f) = \frac{1}{1+G_j}\, \tilde p_j - \frac{G_j}{1+G_j} \frac{\mathscr H_k}{1-\mathscr L_j}\,\tilde h_k. \label{eq:df}
	\end{equation}
This signal now carries not only phase noise but also gravitational wave signals that are imprinted onto the laser phase via the feedback loop.

Plugging the two equations above into equation~(\ref{eq:singlesubtraction}), one can confirm that the subtraction relation still holds. Therefore, the complete recipe~(\ref{eq:postproccess}) is valid even in the presence of the frequency-locked loops.

A beauty of this scheme is that the offline synthesis~(\ref{eq:postproccess}) does not require the full knowledge about the frequency-locked loops. It instead requires an accurate knowledge on the phase responses $ \mathscr L_j$ only. Therefore, the scheme is always applicable regardless of how the frequency-locked loops are implemented as long as they are linear control systems.

A similar idea has been already employed in the ground-based detectors in the context of calibrating the detector output with the main purpose of removing the suppression effect of the interferometer control~\cite{Abbott:2017ki,Cahillane:2017ba}. Indeed, the suppression factor $(1+G_j)^{-1}$ and the open-loop gain $G_j$ completely vanished in our case too as seen in equation~(\ref{eq:singlesubtraction}).

\subsection{Requirements\label{sec:requirements}}
If the simulated cavity responses in the offline data synthesis are inaccurate, they would consequently leave residual phase noise contaminating the synthesized output $X$ due to the incomplete subtraction. We derive the requirements on the accuracies for a set of particular parameters.

For simplicity, a single Fabry-Perot cavity is considered. Replacing the phase response $\mathscr L_j$ by a simulated response $\mathscr L_j^\textrm{(s)}$ in equation~(\ref{eq:singlesubtraction}) where $\mathscr L_j \neq \mathscr L_j^\textrm{(s)}$ and plugging equations~(\ref{eq:yf}) and (\ref{eq:df}) into equation~(\ref{eq:singlesubtraction}), one can find
	\begin{equation}
		\tilde W_j = \left(\mathscr L_j^\textrm{(s)} - \mathscr L_j\right) \frac{\tilde p_j}{ 1+G_j }  + \mathscr H_k \tilde h_k.
	\end{equation}
We have assumed $\mathscr L_j^\textrm{(s)}\approx \mathscr L_j$ for deriving the second term. Comparing the above against the Pound-Drever-Hall signal~(\ref{eq:yf}), one can find that the phase noise is suppressed, in terms of the signal-to-noise ratio, by
 	\begin{equation}
		\textrm{(suppression)} = \frac{\mathscr L_j^\textrm{(s)} - \mathscr L_j}{1 - \mathscr L_j} \left( 1+G_j\right)^{-1}. \label{eq:surpp}
	\end{equation}
The term containing $G$ in the brackets represents the suppression by the frequency-locked loop and is a natural consequence of the fact the BLFP interferometer requires such a control system for keeping the interferometer response linearized and maximized (see Section~\ref{sec:wp}).
    
To achieve a suppression factor of $10^{-8}$, the residual term $(\mathscr L^\textrm{(s)} - \mathscr L)/(1-\mathscr L)$ has to be as small as $10^{-4}$ or smaller, assuming the frequency locked-loop to provide another suppression of $10^{-4}$ in the relevant frequency band. A suppression of $10^{-4}$ from the frequency-locked loop here is a representative value and can be achieved with a moderate control bandwidth of 1~kHz.

According to equation~(\ref{eq:L}), it in turn requires the single-trip time, $T$, and the product of the reflectivities, $r_\textrm{i} r_\textrm{e}$, to be accurately known. Numerically evaluating equation~(\ref{eq:surpp}), we find that the fractional errors for $T$ and $r_\text{i}r_\text{e}$ must be as small as $10^{-4}$ and $10^{-5}$, respectively, for an interferometer with $T= 0.334$~msec and $r_\text{i} =r_\text{e} = 0.9615$.

While these requirements are stringent, they do not seem impractical. For instance, a double-modulation technique demonstrated an accuracy for the absolute length (or equivalently single-trip time) of as low as 70~ppb for Fabry-Perot cavities with lengths of 16~m and 4~km~\cite{Staley:15}, which is sufficient for our requirement on the accuracy for $T$. The product of reflectivities behaves in a similar manner to optical losses which can be characterized at a level of a few~ppm in situ~\cite{Isogai:13}. This also meets the required accuracy for $r_\textrm{i}r_\textrm{e}$. On the other hand, the actual implementations of the measurements in flight have to be carefully thought through.

\subsection{Nonlinearities\label{sec:nonlinear}}
The discussions so far are based on the assumption that all the signals and noises are linear. However, we point out that the response of the Pound-Drever-Hall scheme to the laser frequency is inherently nonlinear. The degree of the nonlinearity increases as the locking point departs from the resonance. Therefore, the locking point must be kept within a certain range around the resonance. In our case, this range is approximately $\pm0.2$ Hz in terms of the laser frequency. If the laser frequency deviates by more than this value, our proposed scheme would not yield a suppression factor of $10^4$ for the subtraction due to the undesired nonlinearity which are not subtracted.

Given our assumption of the pre-stabilization of the laser frequency as explained at the beginning of Section~\ref{sec:offline}, the control bandwidth of 1~kHz is sufficient to keep the locking point within the required range. The residual frequency deviation should be $6\times 10^{-5}$~Hz in rms where a controller gain of $10^4$ and the integration bandwidth of $40$~Hz, corresponding to the cavity pole frequency, are assumed.

The alignment degrees of freedom also couple to the Pound-Drever-Hall readout, forcing the locking point to displace from the resonance point. Similarly to the argument above, misalignment must not push the locking point by more than $\pm0.2$~Hz. Misalignment will produce spurious signals in the Pound-Drever-Hall readout, producing an offset for the operating point. In order for this effect to be negligible, the optical power in TEM01 and TEM10 modes must be smaller than 10\% of that for the TEM00 mode. This corresponds to misalignment by $\sim0.6$ $\mu$rad in the tilt of a cavity mirror and shift in the beam spot by $\sim3$~cm at a cavity mirror where a beam radius of $\sim10$~cm is assumed at the cavity mirrors. These requirement values can be met by implementing an alignment sensing and control system such as those employed in the ground-based detectors.

\section{Signal and noise transfer functions} \label{sec:trans}
\subsection{Response to gravitational waves}
To compute the response of the BLFP interferometer to incoming gravitational waves, we begin with the simple argument of a laser field making a round-trip without a Fabry-Perot cavity. Subsequently, the argument is expanded by adding a Fabry-Perot cavity into consideration.  A portion of the discussions here is based on those described by~\cite{Rakhmanov:2008is}, and therefore we do not fully attempt to  describe the derivations.

\begin{figure}[t]
	\begin{center}
		\includegraphics[width=0.4\columnwidth]{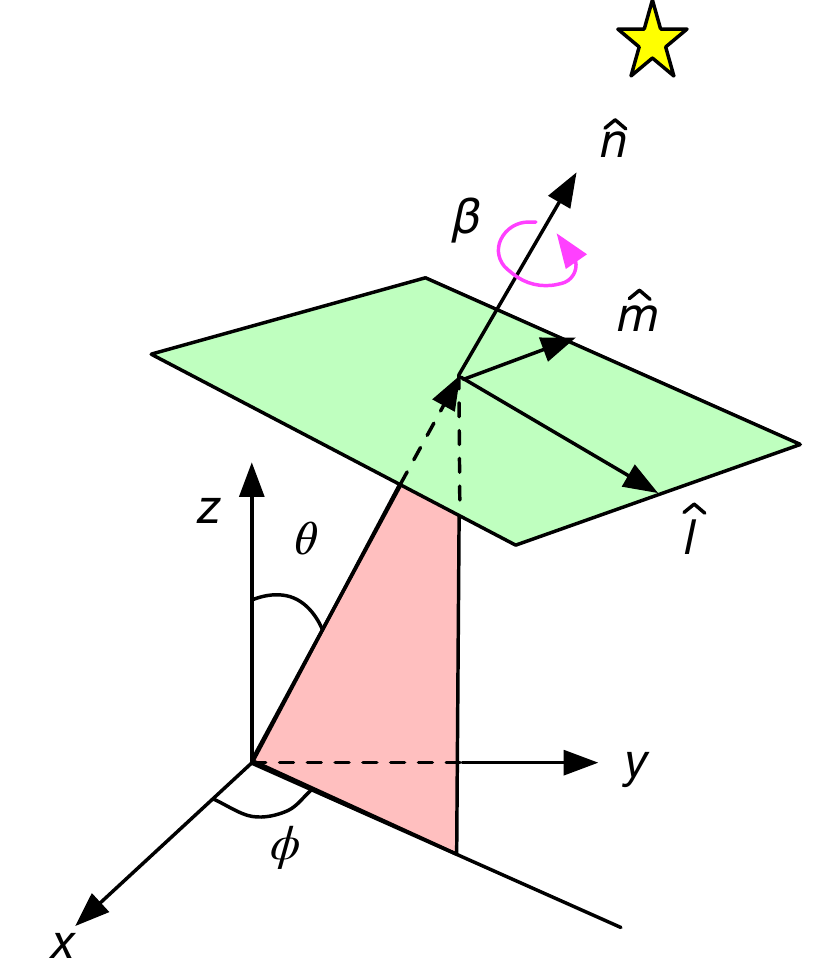}
	\end{center}
\caption{A schematic view of the relevant coordinate systems for computing the gravitational wave response. While the set of unit vectors $(\hat x, \hat y, \hat z)$ forms the observer frame in the right-hand Cartesian, the unit vectors $(\hat l, \hat m, \hat n)$ form the polarization coordinate with an additional rotation along $\hat n$ by $\beta$. \label{fig:coordinates}}
\end{figure}

First of all, our choice of the relevant coordinates is illustrated in Figure~\ref{fig:coordinates}. The observer frame is expressed by the set of unit vectors $(\hat x, \hat y, \hat z)$ forming a right-hand Cartesian coordinate. The BLFP interferometer is assumed to be laid in the $x$-$y$ plane. The other relevant coordinate, called the polarization coordinate~\cite{300}, is given by another set of unit vectors $(\hat l, \hat m, \hat n)$ forming a right-hand Cartesian, too. The vectors $(\hat l, \hat m, \hat n)$ are related to the observer frame via
	\begin{equation}
	\begin{aligned}
		\hat l = &\hat x' \cos \beta + \hat y' \sin \beta, \quad \hat m = -\hat x'\sin\beta + \hat y'\cos \beta\\
		&\quad \textrm{and }\, \hat n = \hat z',  \label{eq:lmn}
	\end{aligned}
	\end{equation}
with $(\hat x', \hat y', \hat z')$ being an intermediate coordinate given by the ordinary coordinate transformation of $(\hat x, \hat y, \hat z)$ through rotations $\theta$ and $\phi$ as indicated in Figure~\ref{fig:coordinates}. They can be explicitly given in terms of the observer coordinate as $\hat x' = (\cos \phi \cos \theta, \sin\phi \cos \theta, -\sin\theta )$, $\hat y' = (-\sin \phi, \cos \phi, 0)$ and $\hat z' = (\cos\phi\sin \theta , \sin \phi \sin \theta, \cos\theta)$. The rotation $\beta$ is often referred to as the polarization angle and corresponds to a rotational degree of freedom of the $l$-$m$ plane about the $\hat n$ axis. For more detailed derivations, see for instance~\cite{Nishizawa:2009ec}. We now define the polarization tensors on the $l$-$m$ plane as
	\begin{eqnarray}
		e^{ij}_+ (\hat n) = l_i l_j - m_i m_j, \label{eq:ep}\\
		e^{ij}_\times (\hat n)= l_i m_j+m_il_j, \label{eq:ex}
	\end{eqnarray}
where the indices $i$ and $j$ run for the spatial coordinates i.e., $i,j = 1, 2, 3$.
Gravitational waves coming from $\hat n$ can be expressed as the sum of the `$+$' and `$\times$' polarization modes,
	\begin{equation}
		h^{ij}(t) = h_+ (t) e^{ij}_+ +  h_\times (t) e^{ij}_\times .
	\end{equation}

If a laser field makes a round trip along a unit vector $\hat a$ in the $x$-$y$ plane, for a single trip distance of $L$, it acquires an additional phase due to the influence of gravitational waves. The size of such phase shift is proportional to the amplitude of the gravitational waves and can be expressed in the frequency domain by
	\begin{equation}
		\tilde \psi ^\textrm{(gw)} (f) = kL a_ia_j D(\hat a, f) \tilde h^{ij}(f),\label{eq:psigw}
	\end{equation}
where $D(\hat a, f)$ describes the frequency-dependent response of the laser field to the gravitational waves, defined by
	\begin{equation}
		D(\hat a, f) = \frac{e^{-2\pi\ii fT }}{2}  \bigg [e^{\ii \pi fT_+} \sinc\left( \pi f T_- \right)+ e^{-\ii \pi fT_-} \sinc\left( \pi f T_+ \right) \bigg ], \label{eq:transD}
	\end{equation}
with short-hand notation: $T_\pm \equiv T\left(1\pm \hat n \cdot \hat a \right)$.

In the case of the Fabry-Perot interferometer, the phase variation (\ref{eq:psigw}) is subsequently amplified by the multiple round-trips, resulting in a phase variation in the leakage field in $E_2$ by
	\begin{equation}
		\tilde \Psi^\textrm{(gw)} (f)= g  \mathscr L \tilde \psi^\textrm{(gw)}(f), \label{eq:Phigw}
	\end{equation}
where $g$ is the cavity gain factor defined by
	\begin{equation}
	g=\frac{1}{1-r_\textrm{i}r_\textrm{e}}.
	\end{equation}
This response is graphically summarized in the block diagram in Figure~\ref{fig:loop} for the Pound-Drever-Hall scheme.

Replacing the place holder $\mathscr H h$ in equation~(\ref{eq:xresult}) by the phase variation~(\ref{eq:Phigw}), one can find the synthesized output for Spacecraft 1 to be
	\begin{equation}
		\tilde X_1 = k\tilde h^{ij} \bigg \{ g_1\left( 1-\mathscr L'_1 \right) \mathscr L_1 L_3a_ia_j D(\hat a) 
		 - g'_1\left( 1-\mathscr L_1 \right) \mathscr L'_1 L_2b_ib_j D(\hat b)   \bigg\}, \label{eq:completeX1}
	\end{equation}
where $\hat b$ is another unit vector aligned to the other Fabry-Perot cavity $L_2$. The two lasers are assumed to have a wavelength almost identical to each other so that $k_1 = k'_1 = k$.  The above completely describes the detector transfer function of incoming gravitational waves $h^{ij}$ to the synthesized output $X_1$.

\begin{table*}[t]
\caption{A list of the noise sources. \label{tab:noise}}

\begin{center}

\begin{tabular}{clc c r}
\hline \hline
&\begin{tabular}{c} {\bf Noise source} \end{tabular}&{\bf Transfer function} & \begin{tabular}{c}{\bf \# of}\\ {\bf sources}\end{tabular} &\begin{tabular}{c} {\bf Input}\\ {\bf magnitude}\end{tabular} \\
 \hline
 (a)&\begin{tabular}{l} Sensing noise for the back\\
 -link heterodyne $\tau_1$ and $\tau'_1$\end{tabular}&  $(1-\mathscr L)^2/2$ & 2 & \begin{tabular}{r} $5.1\times10^{-9}$ \\ rad/$\sqrt{\text{Hz}}$ \end{tabular}\\
 \hline
 (b)&\begin{tabular}{l} Sensing noise for the PDH$^\dagger$ \\
  readouts $y_1$ and $y'_1$\end{tabular}& $1-\mathscr L$ &2 & \begin{tabular}{r} $6.2\times10^{-10}$\\ rad/$\sqrt{\text{Hz}}$ \end{tabular}\\
 \hline
 (c)&\begin{tabular}{l} Residual force noise\\
  on test masses \end{tabular}&  $ k g\mathscr L (1-\mathscr L )/(2 \pi^2  f^2 m) $& 4& \begin{tabular}{r} $1.0\times10^{-16}$\\ N/$\sqrt{\text{Hz}}$ \end{tabular}\\
\hline
(d)&\begin{tabular}{l} Intensity noise of\\  Lasers 1, 1', 2' and 3\end{tabular} &  	$2 k g^2\mathscr L (1-\mathscr L )/(\pi^2c f^2m ) $ & 4& \begin{tabular}{r} $1.4\times 10^{-9}$\\ W/$\sqrt{\text{Hz}}$ \end{tabular}\\
\hline\hline
 
\end{tabular}
\medskip

\textbf{Note--} The labels in the leftmost column are the same as those used in Figure~\ref{fig:NB_Pn}. All the noise inputs is assumed to have a white spectral density across the frequency band. $\dagger$ PDH stands for Pound-Drever-Hall. See Section~\ref{sec:ss} for descriptions on the magnitude of the input noises.

\end{center}
\end{table*}

\subsection{Antenna Pattern}
Based upon the results obtained in the previous subsection, we show that the antenna pattern is almost identical to those for the ground-based detectors except that the amplitude is reduced by a factor of $ \sqrt{3}/2$ for both `$+$'- and `$\times$'-polarization modes. This reduction is a consequence of the fact that the two Fabry-Perot cavities are not orthogonally placed but separated by 60$^\circ$. This is a well known effect~\cite{Schutz:1987fb}.

Starting from the synthesized detector output~(\ref{eq:completeX1}), we assume the two Fabry-Perot cavities to be identical to each other in their properties including the mirror reflectivities and cavity lengths. This reduces the synthesized detector output to
	\begin{equation}
	\begin{aligned}	
		\tilde X_1 = gkL \tilde h^{ij}  \left( 1-\mathscr L \right) \mathscr L \bigg \{ a_ia_j D(\hat a) - b_ib_j D(\hat b)   \bigg\},
	\end{aligned}
	\end{equation}
where $\mathscr L = \mathscr L_1 = \mathscr L'_1$, $L=L_2=L_3$ and $g=g_1=g'_1$.
This can be expressed by the functions called antenna patterns, $H_A$ with $A= +, \times$ as
	\begin{equation}
		\tilde X_1 = 2g kL\left(1-\mathscr L \right) \left[ H_+ \tilde h_+ + H_\times \tilde h_\times \right]. \label{eq:X111}	
	\end{equation}
where,
	\begin{eqnarray}
		H_+(f,\hat n) = \frac{\mathscr L }{2}  \bigg \{ a_ia_j D(\hat a) - b_ib_j D(\hat b)   \bigg\} e^{ij}_+, \label{eq:Hp}\\
		H_\times (f, \hat n )= \frac{\mathscr L }{2}  \bigg \{ a_ia_j D(\hat a) - b_ib_j D(\hat b)   \bigg\} e^{ij}_\times. \label{eq:Hx}
	\end{eqnarray}
Now, bringing these antenna patterns to the long-wavelength limit where $D\rightarrow 1$ and $\mathscr L \rightarrow 1$, one can confirm that they recover the ordinary antenna patterns fully consistent with~\cite{Schutz:1987fb}.

\begin{figure*}[t]
	\begin{center}
		\includegraphics[width=\columnwidth]{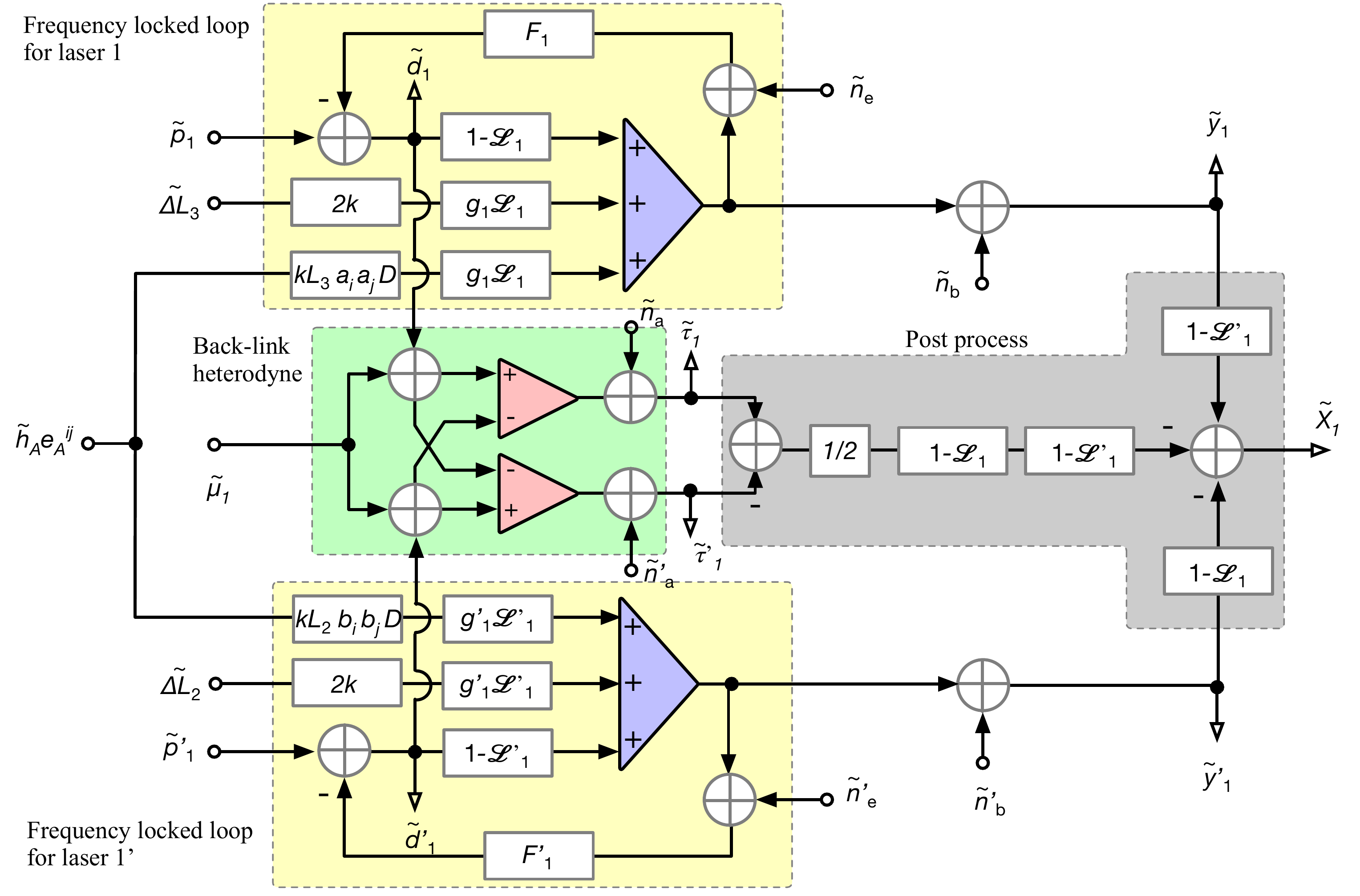}
	\end{center}
\caption{A block diagram representation of the entire system with various noise sources. Each of the two yellow-shaded boxes at the top and bottom represents a frequency-locked loop which has the identical structure to that shown in figure~\ref{fig:loop}. The green-shaded box in the middle represents the back-link heterodyne measurements. The gray-shaded region on the right represents the post process that implements the laser-phase noise subtraction and yields the synthesized output $X_1$.\label{fig:overallloop}}
\end{figure*}

\subsection{Noise transfer functions}
Noises are critical elements in gravitational wave observations because they can hinder the high sensitivity measurements by obscuring gravitational wave signals. In this subsection, we shall evaluate the transfer functions of four major noise sources, including sensing noises associated with the back-link heterodyne measurements, that for the Pound-Drever-Hall readout, residual force noise acting on the test masses and laser intensity noise. A list of the noise sources is given in Table~\ref{tab:noise}.

The most significant noise at high frequencies is identified to be sensing noise in the back-link heterodyne measurements. According to the network diagram illustrated in Figure~\ref{fig:overallloop}, the transfer function to the synthesized output can be characterized as
	\begin{equation}
		\frac{\partial \tilde X_1}{\partial \tilde n_a} = \frac{1}{2}\left( 1 - \mathscr L_1 \right) \left( 1 - \mathscr L'_1\right),
	\end{equation}
with $n_a$ being sensing noise injected at the point right before the $\tau_1$ measurement. The same holds for sensing noise in the other back-link heterodyne $n'_a$, so that $\partial \tilde X_1/\partial \tilde n'_a = \partial \tilde X_1 /\partial \tilde n_a$.

Next, sensing noise in the Pound-Drever-Hall measurement $n_b$ for Laser 1 affects the output via
	\begin{equation}
		\frac{\partial \tilde X_1}{\partial \tilde n_b} = \left( 1-\mathscr L'_1\right).
	\end{equation}
A similar transfer function can be found for sensing noise in the measurement with Laser 1' where the cavity effect $\mathscr L'_1$ is replaced by $\mathscr L_1$.

Residual force and laser intensity noises both come into the measurements by physically displacing the cavity lengths. For residual force noise, the resulting displacement in the $L_3$ cavity can be expressed by
	\begin{equation}
		\widetilde{ \Delta  L_3} = \frac{ \tilde K}{4 \pi^2 f^2 m},
	\end{equation}
where $K$ is residual force noise acting on a test mass given in the unit of Newtons, and where $m$ is the mass of a test mass. 

Similarly, fluctuations in the intensity of the laser circulating in a Fabry-Perot cavity displaces each test mass through the radiation pressure force. The force on a test mass can be given as $2 g^2 t_\textrm{i}^2\Delta P/c$ with $\Delta P$ being fluctuations of the laser power incident on the cavity. Since the radiation pressure force affects the two mirrors coherently, the resulting displacement in the cavity length picks up an additional factor of two and can be expressed for $L_3$ as,
	\begin{equation}
		\widetilde{\Delta  L_3} = \frac{g_1  \widetilde {\Delta P_1} }{ \pi^2 c f^2 m }, \label{eq:int2dl}
	\end{equation}
where we have assumed the cavity to be critically coupled i.e., $r_\textrm{i} = r_\textrm{e}$, leading to $g  = t_\textrm{i}^{-2}$.	

Finally, residual force and laser intensity noises propagate to the synthesized output via the transfer functions, respectively,
	\begin{eqnarray}
		\frac{\partial \tilde X_1}{\partial \tilde K } =  k g_1 \frac{\mathscr L_1\left( 1 - \mathscr L'_1\right)}{2 \pi^2 f^2m},\\
		\frac{\partial \tilde X_1}{\partial  \widetilde{\Delta P}} = 2 k g_1^2 \frac{\mathscr L_1\left( 1 - \mathscr L'_1\right)}{\pi^2 c f^2m}.
	\end{eqnarray}

Additionally, we show that sensing noises in the in-loop Pound-Drever-Hall measurements $n_e$ and $n'_e$ for the frequency-locked loops are subtracted together with laser phase noises. The influence of such sensing noises to the measurements $y_1$ and $d_1$ can be found to be
	\begin{eqnarray}
		\tilde y_1 &=& -\frac{G_1}{1+G_1} \tilde n_e, \\
		\tilde d_1 &=& -\frac{G_1}{1+G_1} \frac{1}{1-\mathscr L_1}\, \tilde n_e.
	\end{eqnarray}
Plugging these into the single-arm version of the data synthesis~(\ref{eq:singlesubtraction}), one can confirm that the right-hand side of the equation becomes zero. 
This holds true for the other frequency-locked loop for Laser 1'. Therefore, from equation~(\ref{eq:X11}), sensing noises for controlling the laser frequencies vanish essentially because they behave in the same manner as laser phase noises do.

\section{Sensitivity Analysis} \label{sec:sens}
Following the prescription detailed in~\cite{Robson:2019iw}, we define the strain spectral density in terms of the square root of the effective noise power spectral density,
	\begin{equation}
		S_n(f) =\frac{P_n(f)}{\mathcal R(f)}, \label{eq:Sn}
	\end{equation}
where $\mathcal R(f)$ is the sky and polarization averaged signal response function of the instrument and $P_n$ is the sum of detector noises under consideration. In this section, we analytically derive $\mathcal R(f)$ and develop $P_n$ by evaluating the contributions from the major noise sources. Finally, we discuss how sensitive a BLFP interferometer may be to the binary coalescence events.

\medskip
\subsection{Sky and polarization averaged response}
Traditionally, the normalized detector output is characterized in relation to incoming gravitational waves via the antenna patterns $H_A$ as
	\begin{equation}
		\tilde s(f) = \sum_{A=+,\,\times} H_A (\theta, \phi, \beta, f) \, \tilde h_A (f).
	\end{equation}
The sky and polarization averaged response $\mathcal R'$ can be defined as
	\begin{equation}
		\big< s^2(f) \big> = \mathcal R'(f) \left\{ \bigg< \left|\tilde h_+\right|^2\bigg> + \bigg< \left| \tilde h_\times \right|^2\bigg> \right\},
	\end{equation}
where $\big< \cdots \big>$ denotes the sky and polarization average,
	\begin{equation}
		\big < x \big> = \frac{1}{4\pi^2} \int_0^\pi d\beta \int_{-1}^1 d(\cos \theta) \int^{2\pi}_0 x\,d\phi.
	\end{equation}
Therefore, the averaged detector response can be related to the antenna pattern functions via
	\begin{equation}
		\mathcal R'(f) = \big< \left| H_+ \right|^2\big> =  \big< \left| H_\times \right|^2\big>.
	\end{equation}
In the case of the BLFP interferometer, using equations~(\ref{eq:Hp}) and (\ref{eq:Hx}), one can find
	\begin{equation}
		\mathcal R'(f) = \frac{3}{20} \left| \mathscr L \right|^2,
	\end{equation} 
for frequencies lower than the cavity free-spectral range or $c/(2L)$. A factor of 3/20 upfront is a consequence of two facts; a factor of 1/5 from the sky and polarization averaging and a factor of $3/4$ from the 60$^\circ$ arm arrangement. Note that this derivation assumes the two arms to be identical to each other in their properties. 

Finally, translating the sky and polarization averaged response $\mathcal R'$ to the one evaluated at the synthesized output $X_1$ via equation~(\ref{eq:X111}), we obtain
	\begin{equation}
	\begin{aligned}
		\mathcal R(f) \equiv& \frac{\left< \left| \tilde X_1 \right|^2\right>}{\left< \left|\tilde h_+\right|^2\right> + \left< \left| \tilde h_\times \right|^2\right>},  \\
		=& \left\{ 2 gk L \left|1-\mathscr L\right|  \right\}^2 \mathcal R'(f).
	\end{aligned}
	\end{equation}
This goes into equation~(\ref{eq:Sn}) when evaluating the strain sensitivity.

\begin{figure}[t]
	\begin{center}
		\includegraphics[width=\columnwidth]{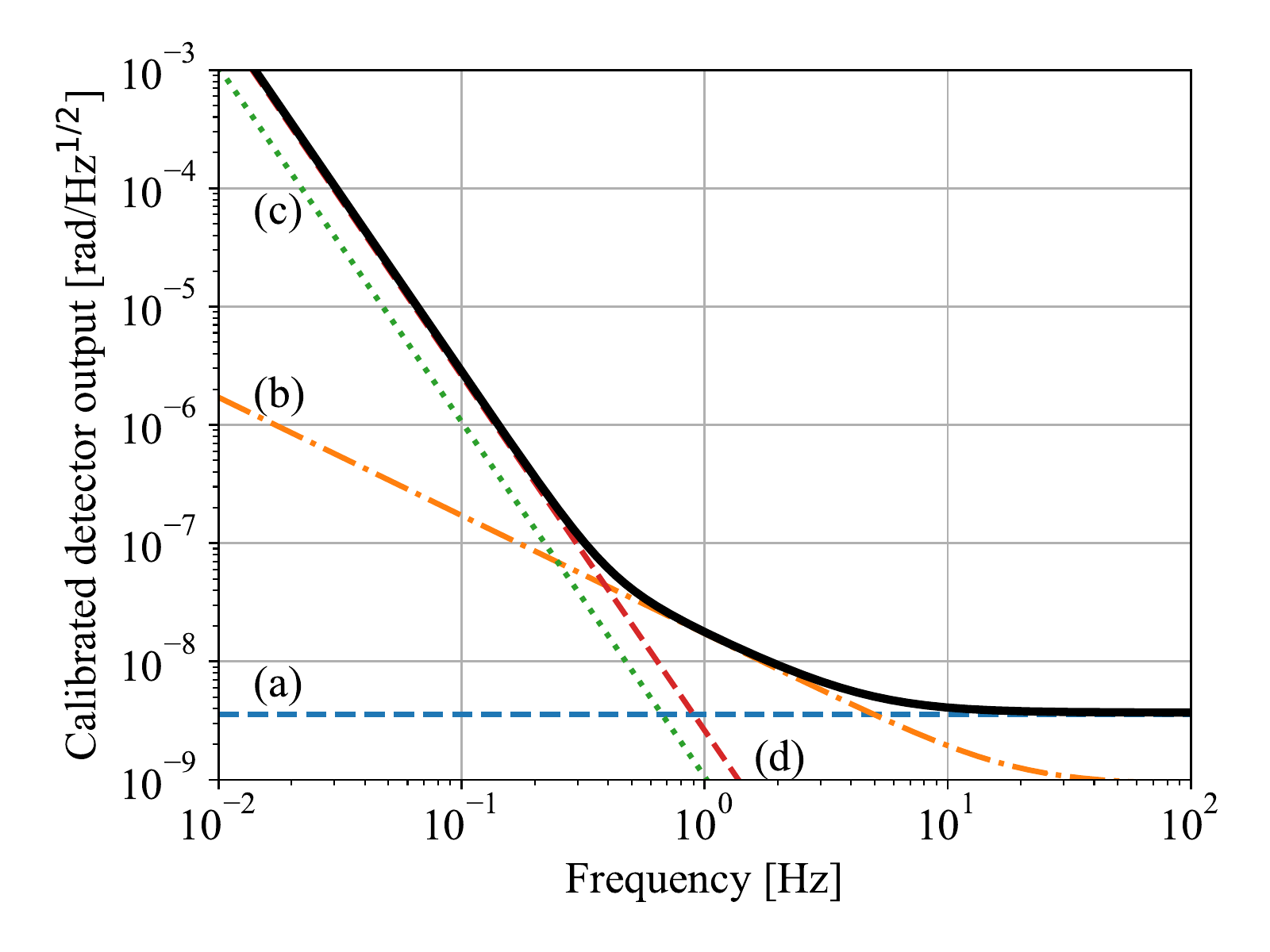}
	\end{center}
\caption{The amplitude spectral density of the synthesized output $\sqrt{P_n}$ (thick black curve) together with the contributions from major noise sources. They are calibrated into the heterodyne optical phase. (a) Sensing noise in the back-link heterodyne measurements. (b) Sensing noise in the Pound-Drever-Hall measurements. (c) Residual force noise. (d) Laser intensity noise. \label{fig:NB_Pn}}
\end{figure}

\subsection{Strain sensitivity}\label{sec:ss}
Prior to evaluating the strain sensitivity, we evaluate the noise power spectrum $P_n$ at the synthesized output $X_1$ by summing up all four major noises described in the previous section. The square root of the resultant power spectral density is shown in Figure~\ref{fig:NB_Pn} together with the contributions from each noise source. The spectra in the figure are calibrated to quantities equivalent to the heterodyne phase in the units of rad/$\sqrt{\textrm{Hz}}$ by multiplying the transfer function $(1-\mathscr L)^{-2}$ to $\sqrt{P_n}$.

The magnitude of each noise source we assumed is summarized in Table~\ref{tab:noise}. As for the interferometer parameters, we adopted the following values. The mass of each test mass is set to $m=30$~kg. The cavity length is set to $L=100$~km. The amplitude reflectivity of both input and end mirrors are set to $r=0.9615$, corresponding to a cavity finesse of 40. The laser wavelength is set to 515~nm and the optical power incident on a side of a Fabry-Perot cavity is set to 500~mW. These values are identical to those proposed for B-DECIGO, a proposed future mission concept~\cite{Nakamura:dl}. 

As shown in Figure~\ref{fig:NB_Pn}, the high frequency band above 5~Hz is limited by sensing noise in the back-link heterodyne measurements. We have assumed sensing noise of $5.1\times 10^{-9} $~rad/$\sqrt{\textrm{Hz}}$ equivalent to optical shot noise for a total optical power of 60~mW in a heterodyne measurement.

The synthesized detector output in the intermediate band from 0.4 to 5~Hz is dominated by sensing noise in the Pound-Drever-Hall measurements. The amplitude spectral density of the noise is assumed to be $6.2\times 10^{-10}$~rad/$\sqrt{\textrm{Hz}}$ corresponding to shot noise with an optical power of 500~mW incident on each Fabry-Perot cavity. We have assumed almost all the optical power in reflection is directed to the $y_1$ ($y'_1$) measurement for minimizing the noise contribution.

\begin{figure}[t]
	\begin{center}
		\includegraphics[width=\columnwidth]{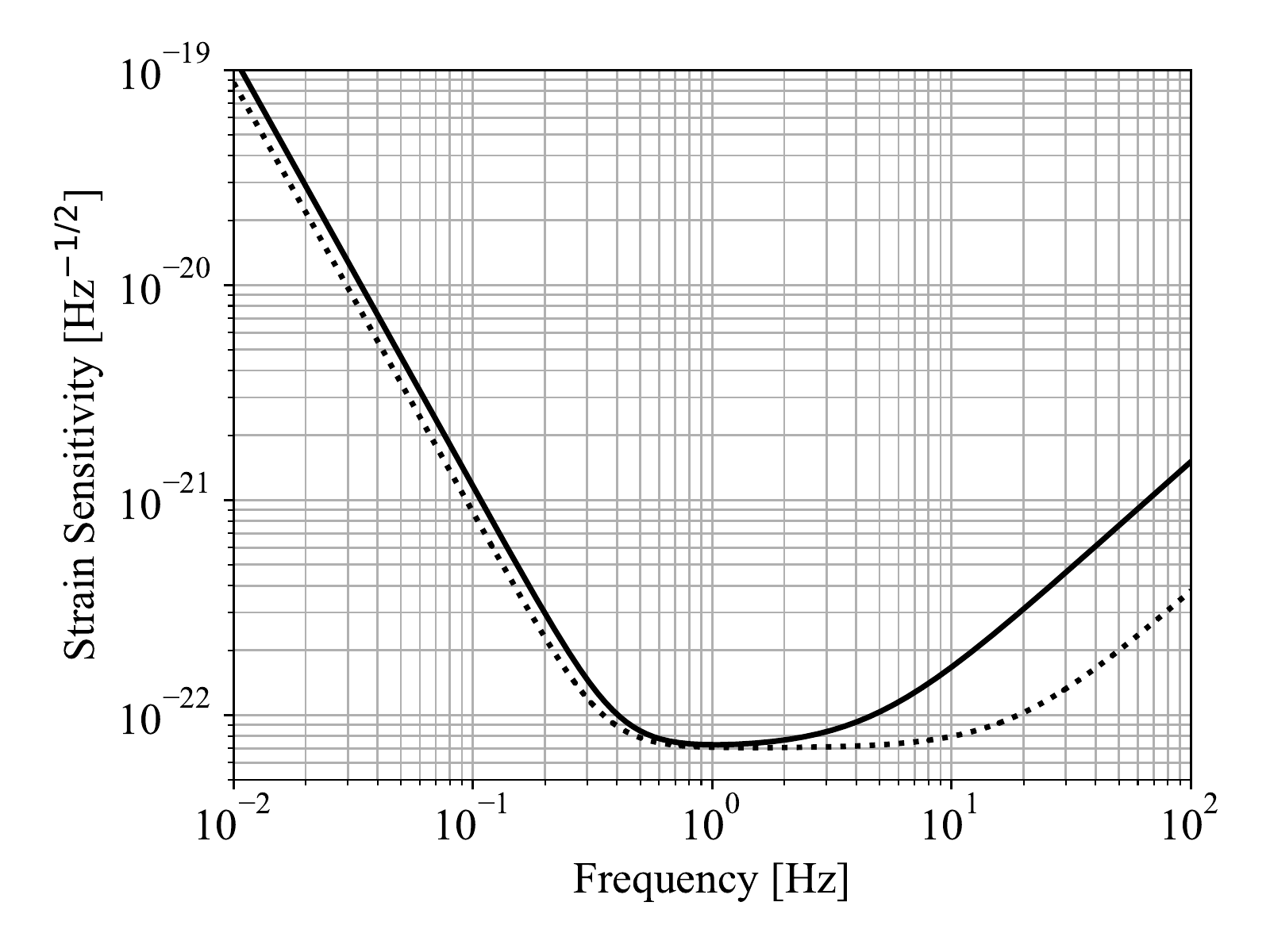}
	\end{center}
\caption{The amplitude spectral density of the sky- and polarization averaged strain sensitivity $\sqrt{S_n}$ for the BLFP interferometer (shown as solid curve) in comparison to the conventional interferometer without the BLFP topology (shown as dotted curve). The sensitivities are for a two-arm interferometer and does not include the contributions from the other two interferometers that are available out of a single constellation. \label{fig:sensitivity}}
\end{figure}

The rest of the low frequency band below 0.4~Hz is found to be limited chiefly by laser intensity noise. An ambitious intensity stability of the input lasers with relative intensity noise of $2.8\times 10^{-9}$~/$\sqrt{\textrm{Hz}}$ across the frequency band is assumed.
This means that the laser intensity has to be stabilized with an optical power monitor system limited by photon shot noise for an optical power of 100~mW. The noise contribution also includes the ones from the lasers onboard the far spacecraft. For instance, Laser 2' onboard Spacecraft 2 displaces the length of $L_3$ through the radiation pressure force. Consequently, the resultant contributions from laser intensity noise becomes predominant over residual force noise. The magnitude of residual force noise is assumed to be $1.0\times 10^{-16} $~N/$\sqrt{\text{Hz}}$ per test mass. The residual force noise is assumed to be dominated by Brownian force noise from residual gas surrounding the test masses at a 0.01~$\mu$Pa level, which demands a more stringent requirement for the vacuum level than LISA by a factor of 100.

Finally, plugging the resulting $P_n$ and the averaged response $\mathcal R$ into equation~(\ref{eq:Sn}), we obtain the strain sensitivity spectrum of the BLFP interferometer as shown in Figure~\ref{fig:sensitivity}. The sensitivity reaches approximately $7\times 10^{-23}$~$\text{Hz}^{-1/2}$ in the band at around 1 Hz while it degrades below 0.4~Hz as $f^{-2}$ and above 5~Hz as $f$, due to the noise contributions as described above.

For a fair comparison, we also evaluated the sensitivity of the conventional interferometer without employing the BLFP topology while keeping the same detector parameters, as shown in Figure~\ref{fig:sensitivity}. The contamination by laser phase noise in the conventional interferometer is neglected by assuming a perfect common-mode noise rejection. It is clear that the BLFP interferometer introduces one additional noise, namely sensing noise associated with the back-link heterodyne measurements. This particular noise contaminates the strain sensitivity via a transfer function proportional to $f$ and hence the degraded sensitivity at high frequencies. In addition, the BLFP interferometer suffers from an increased noise contribution from laser intensity noise by a factor of $\sqrt{2}$ at low frequencies due to the fact that there is no common-mode rejection between Lasers 1 and 1'. Such increased noises are the prices that one has to pay for obtaining the technical advantages of the BLFP interferometer.

\begin{figure}[t]
	\begin{center}
		\includegraphics[width=\columnwidth]{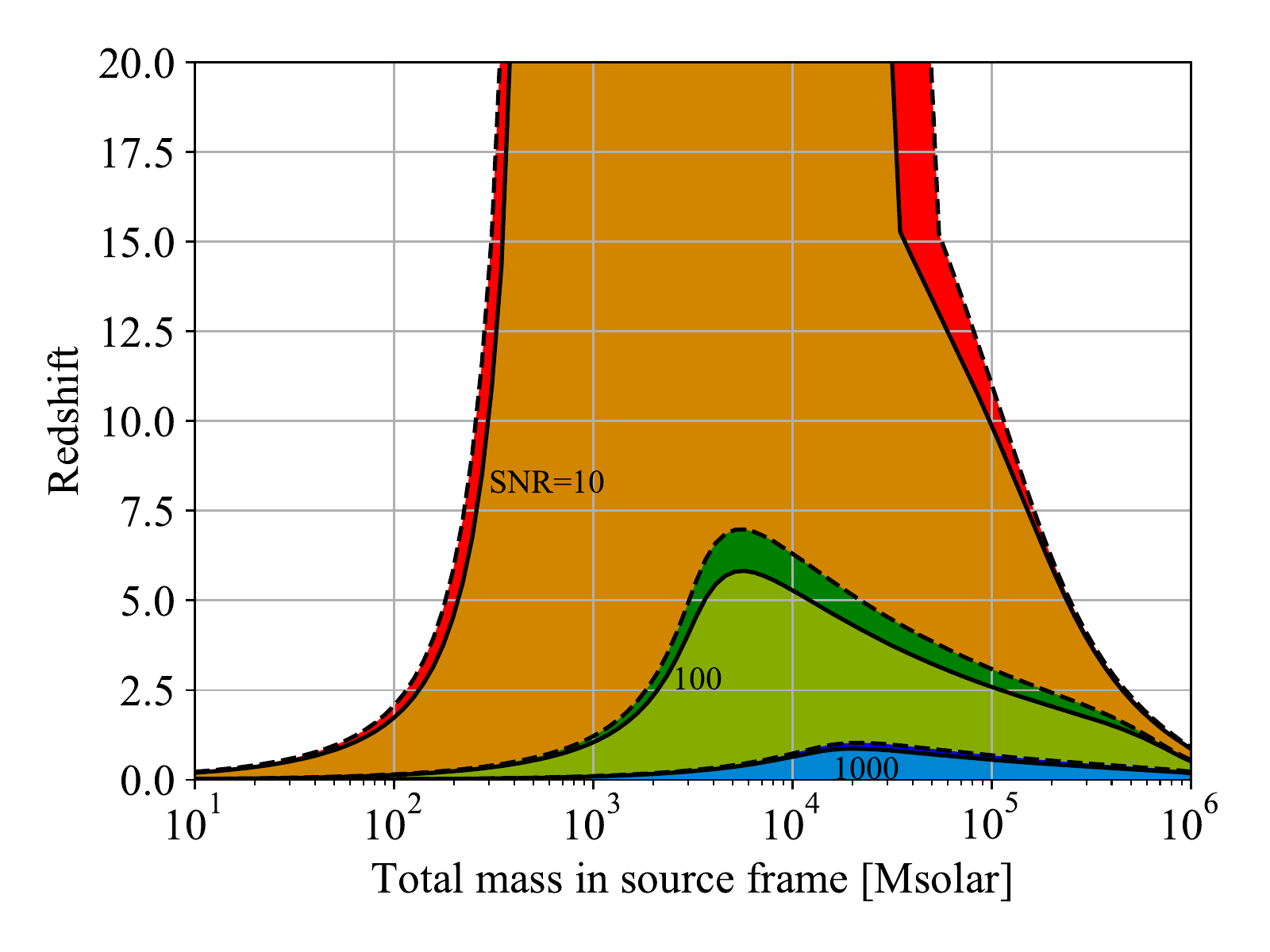}
	\end{center}
\caption{Binary detection ranges as functions of the total source-frame mass of binary coalescences, given a few amplitude signal-to-noise ratios, including $\rho = 10$, 100, and 1000. Those drawn by dashed lines are for the interferometer without the BLFP configuration. The computation assumed the standard $\Lambda$CDM cosmology model with the 2015 Planck results plugged~\cite{Planck2015}.\label{fig:binary}}
\end{figure}

\subsection{Implications for the binary detection range}
A figure of merit for assessing the observatory performance is often given in the form of the binary detection range. The binary detection range indicates redshifts or luminosity distances to which the observatory can detect compact binary coalescences for a given threshold signal-to-noise ratio.

Figure~\ref{fig:binary} shows a few examples of the binary detection range for the BLFP interferometer for amplitude signal-to-noise ratios of $\rho = 10, 100$ and 1000. We followed the computation detailed in~\cite{Robson:2019iw} for relating the signal-to-noise ratio to the sky and polarization averaged strain sensitivity. The strain sensitivity of the BLFP is assumed to be the one shown in Figure~\ref{fig:sensitivity}. Phenom A~\cite{Ajith_2007,PhysRevD.77.104017} was used as the waveform of binary coalescences. The binary systems are assumed to have a constant mass ratio of $m_2/m_1 = 0.2$.

For an amplitude signal-to-noise ratio of 10, the BLFP interferometer is able to reach redshift of 10 or greater for the binary systems with the total mass from a few $10^2$~$M_\odot$ to $10^5$~$M_\odot$. Therefore, it can observe essentially almost all the binary systems in our Universe in the mass range corresponding to the intermediate-mass black holes.

Moreover, events similar to GW150914~\cite{Abbott:2016ki} with the total mass of approximately 60~$M_\odot$ at a redshift of $z=0.09$ or luminosity distance of $D_L = $410~Mpc are found to be well within the detection range. This means that the BLFP interferometer is capable of contributing to multi-wavelength observations of such events in collaboration wth the ground-based detectors for improving the ability of accurately estimating the astrophysical parameters of the binary systems. Additionally, a forewarning to the ground-based detectors for the events can be issued such that they can be prepared a day or some hours before the merger.

For comparison, the binary detection ranges for the conventional interferometer without the BLFP is overlaid in Figure~\ref{fig:binary}. The conventional one exhibits better detection ranges at all binary masses because of the better sensitivity at all frequencies. We find the discrepancy in the detection range to be only 10-20~\%  for the case for a signal-to-noise ratio of 100 as shown in Figure~\ref{fig:binary}. Therefore, the improved technological feasibility, as discussed in Section~\ref{sec:adv}, comes with such compromise in the astrophysical performance. This is a trade-off point that the mission designers have to carefully examine once the science goals are determined.

\medskip
\section{Discussions}\label{sec:discussion}
Even though the BLFP configuration is expected to increase the feasibility of space laser interferometry equipped with the Fabry-Perot cavities, it requires several technological developments and stuides.

One is the phasemeter. As shown in Figure~\ref{fig:NB_Pn}, the back-link heterodyne measurements must be able to resolve phase variations at a level of $10^{-9}-10^{-8}$~rad/$\sqrt{\textrm{Hz}}$. Therefore, the phasemeter extracting the optical phase must have internal noise lower than that in order not to contaminate the strain sensitivity. The phasemeter based on a digital phase-locked loop marks a phase precision of $10^{-6}$ rad/$\sqrt{\textrm{Hz}}$ at frequencies relevant to the BLFP~\cite{doi:10.1063/1.2405113}. Another study showed an excellent precision of $5\times 10^{-8}$ rad/$\sqrt{\textrm{Hz}}$ with the use of a zero-crossing algorithm~\cite{Kokuyama_2016}. In either implementation, some advancement is necessary.

Another point is associated with the calibration of the Fabry-Perot response $\mathscr L$ as described in Section~\ref{sec:requirements}. The calibration is critical because inaccurate knowledge on them will directly lead to incomplete noise subtraction. Unlike time-delay interferometry where only the time delay needs to be accurately determined, our scheme needs two quantities to be accurately determined for each Fabry-Perot cavity i.e., the single trip time $T$ and the product of the mirror reflectivities $r_\text{i} r_\text{e}$ according to equation~(\ref{eq:L}). Some technical assessment should be made to derive requirements on their accuracies and to study practical limits. Possible solutions include the Wiener filtering technique as was implemented in the ground-based detectors~\cite{Driggers:2019gx}. If necessary, temporal variations of the responses have to be monitored and compensated~\cite{Tuyenbayev:2016ka}.

The photo-detection system in the back-link heterodyne measurements must be carefully designed to be able to handle a high optical power.  The photodetectors exposed to high optical power can easily run into the saturation issue. This point was raised in the conceptual design of BBO~\cite{Harry:2006fz}. A straightforward mitigation is the use of multiple photodetectors so as to maintain the optical power on the individual photodetectors as low as necessary.

Finally, thorough study must be conducted to cope with the frequency pulling effect~\cite{PhysRevD.90.062005} due to the orbital motions of the spacecraft. Assuming an orbital design similar to LISA~\cite{Dhurandhar_2005,Nayak_2006} and scaling the relative spacecraft motions for a length of 100~km, we foresee slow variations in the frequencies of the locked lasers to be approximately 100~Hz/sec. If the hardware is able to handle frequency drifts up to a few tens of MHz, the duration for a continuous observation period would be limited to a few days. This may indicate that the lasers have to be relocked to different resonance points of the Fabry-Perot cavities to reset the accumulated frequency drift on a time scale of a few days. In any case, a robust observation strategy must be developed.

\section{Conclusions}\label{sec:conclusion}
A new topology for laser interferometer, named the back-linked Fabry-Perot interferometer, has been proposed for probing gravitational waves in space in the deci-Hz band with the goal of fully exploiting the potential of gravitational wave astronomy. The back-linked Fabry-Perot interferometer is able to keep all the Fabry-Perot cavities at resonances by employing the set of frequency-locked loops while circumventing the noise contamination from laser phase noises via a new noise subtraction scheme. Therefore, it can remove the need for the high-precision control of the cavity lengths and simultaneously alleviate the requirement on laser frequency stabilization. These practical advantages should lead to mission designs which are technologically more feasible.

This article, for the first time, presented the working principle of the back-linked Fabry-Perot interferometer and the idea of phase-noise subtraction dedicated for it. The transfer functions for gravitational waves and the major noise sources are carefully derived as part of the sensitivity analyses. We figured that the strain sensitivity at high frequencies above 5~Hz could be contaminated by sensing noise in the back-link heterodyne measurements that were unique to the back-linked Fabry-Perot interferometer. Nonetheless, our analysis suggested that a strain sensitivity of $7\times10^{-23}\, \textrm{Hz}^{-1/2}$ at around 1~Hz was achievable, given a set of moderately ambitious parameters. This corresponds to an observatory performance being able to probe binary systems farther than $z=10$ for the total mass of $10^2-10^5$~$M_\odot$.

Among several technological challenges, the phase noise subtraction has been identified to be the most critical because imperfect subtraction would significantly degrade the observatory performance. In order to assess the feasibility of the subtraction scheme, we are currently planning to perform an experimental verification in laboratory.
We conclude that the back-linked Fabry-Perot interferometer can potentially be a candidate for future space gravitational wave missions once the technological challenges are fully addressed.

\section*{Acknowledgments}
The authors are grateful to M.~Ando, Y.~Michimura and K.~Nagano of the University of Tokyo for fruitful and stimulating discussions. R.~Sugimoto contributed to this work by providing crosscheck for a part in the binary range computation. This work was supported by JSPS KAKENHI Grant Number JP20H01938.

\appendix
\medskip
\section{Derivation of the Fabry-Perot response}\label{sec:pdh}
We derive equation~(\ref{eq:ytfp}) for the specific case where the Pound-Drever-Hall readout scheme is employed for a Fabry-Perot cavity. The electric field of the laser incident on the Fabry-Perot cavity has to be phase-modulated by a monochromatic excitation and, therefore, can be expressed by
	\begin{equation}
		E_\text{in}(t) = A e^{\ii \Omega t + \ii p(t) } \exp\left( \ii \Gamma \sin \omega_\textrm{m} t\right),
	\end{equation}
where $\Omega$, $\omega_\textrm{m}$ and $\Gamma$ are the laser's angular frequency, the angular frequency of the excitation and the modulation depth, respectively.
Using the Jacobi-Angar relation, one can expand the above into the collection of three distinct laser fields as
	\begin{equation}
		E_\text{in} = A e^{\ii \Omega t + \ii p(t) }\left\{ J_0(\Gamma) + J_1(\Gamma) e^{\ii \omega_\textrm{m} t} - J_1(\Gamma) e^{-\ii \omega_\textrm{m} t}\right\},
	\end{equation}
where $J_n$ is the Bessel function of the first kind.
The first term represents the carrier field while the second and third are the upper and lower sidebands, respectively. 

Since each field has a different laser frequency i.e., $\Omega$ and $\Omega \pm \omega_\textrm{m}$, they will acquire different reflection coefficients when they are reflected off of the cavity. The laser fields in reflection can be generally given as,
	\begin{equation}
		E_\textrm{refl} (t)= e^{\ii \Omega t} \left[E^\textrm{(c)} + E^\textrm{(u)} e^{\ii \omega_\textrm{m} t}+E^\textrm{(l)} e^{-\ii \omega_\textrm{m} t} \right],
	\end{equation}
where $E^\textrm{(a)}$ for $\textrm{a}= (\textrm{c, u, l})$ are complex field amplitudes.

We now look into its optical power in reflection because that is the quantity one can measure in reality. Extracting the terms oscillating at frequencies near $\omega_\textrm{m}$, one can find 
	\begin{equation}
	\begin{aligned}
		P^{(\omega_\textrm{m})} &= -2\Im\left[ \left(E^\textrm{(l)}\right)^\dagger E^\textrm{(c)} + \left( E^\textrm{(c)} \right)^\dagger E^\textrm{(u)} \right] \sin \omega_\textrm{m} t
		\\ &+ 2 \Re \left[ \left(E^\textrm{(l)}\right)^\dagger E^\textrm{(c)} + \left( E^\textrm{(c)} \right)^\dagger E^\textrm{(u)} \right] \cos \omega_\textrm{m} t, \label{eq:Pmod}
	\end{aligned}
	\end{equation}
where a $\dagger$ sign denotes the complex conjugate.	
The Pound-Drever-Hall scheme extracts the term in the first square brackets (often called the in-phase term) by electronically demodulating the above at $\omega_\textrm{m} $.

Now, let us evaluate the cavity reflectivity by virtually letting the following monochromatic field be incident on the cavity,
	\begin{equation}
		E'_\text{in}(t) = A'\exp\left\{ \ii \Omega' t + \ii p(t) \right\}.
	\end{equation}	
The leakage field in reflection (see Figure~\ref{fig:virtualFP}) can be expressed by the summation of the fields which experienced $n$ round trips as
	\begin{equation}
		E_\text{leak} =A' t_i^2 r_e e^{\ii\Omega' t}\sum_{n=1}^\infty e^{\ii p(t-2nT)}e^{-2n\ii \Omega' T}  \left( r_\text{i} r_\text{e} \right)^{n-1}.
	\end{equation}
Assuming the phase deviation $p(t)$ to be sufficiently small so that $e^{\ii p(t)} \approx 1+ \ii p(t)$ and plugging the Fourier transform~(\ref{eq:fourier}), one obtains
	\begin{equation}
	\begin{aligned}
		E_\text{leak}(t)& = A' e^{\ii\Omega' t}\bigg\{ \frac{t_\text{i}^2 r_\text{e} e^{-2\ii \Phi}}{1 - r_\text{i} r_\text{e}e^{-2\ii \Phi }} + \ii  t_\text{i}^2 r_\text{e} \int^\infty_{-\infty} \tilde p(\omega) \frac{e^{-2\ii \omega T -2 \ii \Phi }}{1 -r_\text{i}r_\text{e} e^{-2\ii  \omega T -2\ii \Phi}} e^{\ii\omega t} d\omega \bigg\}, \label{eq:leak}
	\end{aligned}
	\end{equation}
where we have introduced the single-trip optical phase $\Phi\equiv \Omega' T$.

For the carrier field, since it is resonant i.e., $\Phi = \pi l$ with $l$ being an arbitrary integer, the leakage field~(\ref{eq:leak}) becomes
		\begin{equation}
		E^\textrm{(c)}_\textrm{leak} (t) = A J_0e^{\ii\Omega t}\frac{t_\text{i}^2 r_\text{e} }{1 - r_\text{i} r_\text{e}} \left\{ 1+ \ii  \ell \ast p \right\}. \label{eq:phaseres2}
	\end{equation}
It is handy to rewrite the above by converting the phase term back into the exponential format such that
	\begin{equation}
		E^\textrm{(c)}_\textrm{leak}  (t) = g' AJ_0  e^{\ii \Omega t}  e^{\ii \ell \ast p},
	\end{equation}
where $g'$ is a gain factor defined by $g' = t_\text{i}^2 r_\text{e} /(1 -r_\text{i} r_\text{e})$. As illustrated in Figure~\ref{fig:virtualFP}, the reflected field is composed by the sum of the leakage and prompt-reflection fields. Thus, the complex amplitude of the carrier field in reflection can be given as
	\begin{equation}
		E^\textrm{(c)} (t)= A J_0   \left\{ -r_\textrm{i}e^{\ii p(t)} +  g' e^{\ii \ell \ast p} \right\}. \label{eq:c}
	\end{equation}

Similarly, the complex amplitude of the upper sideband field in reflection can be derived as
	\begin{equation}
		E^\textrm{(u)} (t) =\hat r A J_1   e^{\ii p(t)}, \label{eq:u}
	\end{equation}
where we have assumed the upper sideband field to be almost anti-resonant with the cavity i.e., $\Phi = \pi(1/2 +l)$ and where $\hat r$ is the amplitude reflectivity for the anti-resonant condition, defined by $\hat r \equiv -(r_\textrm{i} + (r_\textrm{i}^2 + t_\textrm{i}^2)r_\textrm{e})/(1+r_\textrm{i}r_\textrm{e})$. We have dropped off the term corresponding to the second term in equation~(\ref{eq:leak}) because it is negligible. Similarly, the lower sideband field in reflection is derived to be
	\begin{equation}
		E^\textrm{(l)} (t) =-\hat r AJ_1   e^{\ii p(t)}, \label{eq:l}
	\end{equation}
where we have assumed the lower sideband field to be almost anti-resonant with the cavity, too.

Finally, plugging equations~(\ref{eq:c}), (\ref{eq:u}) and (\ref{eq:l}) into the optical power~(\ref{eq:Pmod}), one can find 
	\begin{equation}
		P^{(\omega_\textrm{m})} = -4 \hat r J_0J_1A^2g'\left( p(t) - \ell \ast p \right)  \sin \omega_\textrm{m} t,
	\end{equation}
where we converted the terms associated with the phase deviation back to the first-order expansion. The Pound-Drever-Hall scheme then demodulates this signal at $ \omega_\textrm{m}$, which essentially removes $\sin \omega_\textrm{m}t$ from the above. Finally, normalizing the above into an equivalent to the optical phase, we arrive at equation~(\ref{eq:ytfp}). 

\section{Characteristics of the Fabry-Perot response}\label{sec:char}
Let us review a couple of characteristics of the Fabry-Perot response~(\ref{eq:ytfp}). We start from approximating the linear operator~(\ref{eq:L})  to a single pole representation as
	\begin{equation}
		\mathscr L(f) \approx \frac{1}{1+\ii f/f_c}, \label{eq:Lapp}
	\end{equation}
where $f_c \equiv (1-r_\textrm{i}r_\textrm{e})/(4\pi T)$ represents the cut-off frequency called the cavity pole. This is merely a first-order low-pass filter with a cut-off at $f_c$. Plugging it into the response in the frequency domain~(\ref{eq:yfd}), one can find it to be $\tilde y = \ii \tilde p  ff_c^{-1}(1+\ii f/f_c)^{-1}$. Since the laser phase is interconvertible to the laser frequency via the relation $\ii f \tilde p = \tilde \nu$ with $\nu$ being the laser frequency, the response can be rewritten as,
	\begin{equation}
		\tilde y (f) = \frac{1}{1+\ii f/f_c} \frac{ \tilde \nu (f)}{f_c}.
	\end{equation}
This is consistent with the common understanding that the Fabry-Perot cavity is a linear sensor for the laser frequency~\cite{doi:10.1119/1.1286663} with a cut off at $f_c$.

Additionally, the single pole approximation can be further approximated to merely a time-delay in the low-frequency limit or $f\ll f_c$. Imposing such limit on equation~(\ref{eq:Lapp}), we obtain
	\begin{equation}
		\mathscr L(f) \approx e^{-\ii f /f_c} \quad (\textrm{for } f\ll f_c), \label{eq:simpledt}
	\end{equation}
which is equivalent to the frequency domain representation of a time delay $\mathcal D (\tau_c)$ with $\tau_c$ being $\tau_c = 2 T/(1-r_\textrm{i}r_\text{e})$. Therefore, the Fabry-Perot response~(\ref{eq:yfd}) becomes identical to the Michelson response~(\ref{eq:ymi}) in the low-frequency limit. Since the cavity is typically designed to have a high finesse i.e., $(1-r_\text{i}r_\text{e})^{-1} > 1$, the round trip time $2T$ is effectively increased. This is precisely the reason that the Fabry-Perot cavity enhances the phase sensitivity as described in many literatures elsewhere, for example in~\cite{Saulson}. We emphasize that we intend to use equations~(\ref{eq:ytfp}) and (\ref{eq:L}) for the offline noise subtraction as opposed to the simple time delay~(\ref{eq:simpledt}).

\bigskip

\vspace{0.2cm}
\noindent
\let\doi\relax



\end{document}